\documentclass[12pt]{JHEP3}
\input epsf.tex
\epsfclipon
\let\ifpdf\relax

\usepackage{epsfig}
\usepackage{epstopdf}
\usepackage{epsf}
\input{epsf.sty}
\usepackage{graphicx,amsmath,amssymb}
\usepackage{epsfig,multicol}
\allowdisplaybreaks

\usepackage{bbm,bm,amsmath,amssymb}
\def\blfootnote{\xdef\@thefnmark{}\@footnotetext}

\long\def\symbolfootnote[#1]#2{\begingroup%
\def\thefootnote{\fnsymbol{footnote}}\footnote[#1]{#2}\endgroup}

\newcommand{\be}{\begin{eqnarray}}
\newcommand{\ee}{\end{eqnarray}}
\newcommand{\ben}{\begin{eqnarray*}}
\newcommand{\een}{\end{eqnarray*}}

\newcommand{\bcent}{\begin{center}}
\newcommand{\ecent}{\end{center}}
\newcommand{\benum}{\begin{enumerate}}
\newcommand{\eenum}{\end{enumerate}}
\newcommand{\bdesc}{\begin{description}}
\newcommand{\edesc}{\end{description}}
\newcommand{\bitem}{\begin{itemize}}
\newcommand{\eitem}{\end{itemize}}
\newcommand{\bquote}{\begin{quote}}
\newcommand{\equote}{\end{quote}}
\newcommand{\bhalfp}{\begin{minipage}{0.45\textwidth}}
\newcommand{\ehalfp}{\end{minipage}}
\newcommand{\bhead}{\begin{center}\bf \Large}
\newcommand{\ehead}{\end{center}\bigskip}

 \newcommand{\tildeC}{{\tilde{C}}}

\def\be{\begin{equation}}
\def\ee{\end{equation}}
\def\ba{\begin{eqnarray}}
\def\ea{\end{eqnarray}}

\newcommand{\roughly}[1]{\mathrel{\raise.3ex\hbox{$#1$\kern-0.85em
\lower1ex\hbox{$\sim$}}}}

\def\2pi{\left(2\pi\right)}

\def\beq{\begin{equation}}
\def\eeq{\end{equation}}
\def\bg{\begin{eqnarray}}
\def\nd{\end{eqnarray}}
\def\bea{\begin{eqnarray}}
\def\eea{\end{eqnarray}}

\def\D3{\overline{\mbox{D3}}}

\title{de Sitter Vacua in Type IIB String Theory: Classical Solutions and Quantum Corrections}

\author{Keshav Dasgupta${}^1$, Rhiannon Gwyn${}^2$, Evan McDonough${}^1$, Mohammed Mia${}^3$, Radu Tatar${}^4$\\
\vskip.03in
${}^1$ Ernest Rutherford Physics Building, McGill University,\\
3600 University Street, Montr{\'e}al QC, Canada H3A 2T8\\
{\tt evanmc@physics.mcgill.ca, keshav@hep.physics.mcgill.ca}\\
${}^2$ AEI Max-Planck-Institut f\"ur Gravitationsphysik,\\
D-14476 Potsdam, Germany\\
{\tt rhiannon.gwyn@aei.mpg.de}\\
${}^3$ Department of Physics, Purdue University,\\
525 Northwestern Avenue, W. Lafayette, IN 47907-2036\\
{\tt mmia@purdue.edu}\\
${}^4$ Department of Mathematical Sciences,\\
The University of Liverpool, Liverpool, L69 3BX, England, U.K\\
{\tt Radu.Tatar@Liverpool.ac.uk}}

\abstract{We revisit the classical theory of ten-dimensional two-derivative gravity coupled
to fluxes, scalar fields, D-branes, anti D-branes and Orientifold-planes. We show that such 
set-ups do not give rise to a four-dimensional positive curvature spacetime with
the isometries of de Sitter spacetime. We further argue
that a de Sitter solution in type IIB theory may still be achieved if the higher-order curvature corrections are carefully controlled. Our analysis relies on the
derivation of the de Sitter condition from an explicit background solution by
going beyond the supergravity limit of type IIB theory. As such this also tells
us how the background supersymmetry should be broken and under what conditions
D-term uplifting can be realized with non self-dual fluxes.}

\maketitle

\begin{document}


\section{Introduction}

The hot, dense state of the early universe and its subsequent evolution offer a unique testing ground for theories of high-energy physics; if string theory is the correct theory of the earliest universe, it should be possible to embed all the known results from cosmology in a consistent string theory description. Our best observational data of the early universe, from the cosmic microwave background (CMB)\cite{Smoot:1992td, Komatsu:2010fb, Dunkley:2010ge, Story:2012wx, Ade:2013uln, Ade:2013zuv}, and late time acceleration  \cite{Weinberg:2012es}, point to a universe that is very close to spatially flat, in which large-scale structure was generated from an almost scale-invariant spectrum of primordial density perturbations with a nearly Gaussian distribution. This is consistent with a large class of inflationary models \cite{Ade:2013uln}, which we will have in mind here, as well as a variety of alternatives to inflation \cite{Brandenberger:2011gk, Lehners:2008vx,Lehners:2013cka}.

However, the dynamics of the early universe is necessarily studied via an effective field theory (EFT) approach.  Although one might expect a decoupling of energy scales, leading to suppression of higher-order terms in the Lagrangian by increasing powers of the cut-off, the predictions of inflation can be highly sensitive to corrections of both the potential or inflaton mass \cite{Baumann:2009ni} and the kinetic terms \cite{Silverstein:2003hf, Franche:2009gk}.  This forces one to consider the UV sensitivity of inflation, which has been addressed from many perspectives: see \cite{Baumann:2009ni, Silverstein:2013wua} for reviews,  \cite{Burgess:2014lza} for a recent take, and  \cite{Martin:2000xs} for a completely different approach. The dependence of cosmological observables on the detailed embedding of inflation into string theory offers a unique window into the high-energy physics of the early universe, and may provide evidence that string theory could be the correct description of physics at these scales.

A consistent string compactification with a de Sitter (or quasi de Sitter) vacuum in the 3+1 non-compact directions is crucial to such an embedding. Achieving such a compactification has proved to be an extremely difficult endeavour. No-go theorems exist for supergravity \cite{Gibbons:1984kp, Gibbons:2003gb} and for string theory (without time-dependent fields or higher-curvature corrections), the well-known Maldacena-Nunez result \cite{Maldacena:2000mw}. This was extended to the heterotic case with higher-order corrections (but without non-perturbative effects) included \cite{Green:2011cn, Gautason:2012tb}.

In Type II string theory, dS solutions have been studied in many works, for example \cite{Hertzberg:2007wc, Haque:2008jz, Covi:2008ea, Maloney:2002rr, Douglas:2010rt, Neupane:2009jn, Neupane:2010is, Cicoli:2012vw, Cicoli:2012fh, Cicoli:2013mpa, Cicoli:2013cha}. In addition, many models of inflation in string theory have been proposed (see the reviews by \cite{Silverstein:2013wua, Baumann:2009ni}), together with `uplift' mechanisms for obtaining dS \cite{Kachru:2003aw, Burgess:2003ic, Westphal:2006tn} by lifting an AdS minimum of the scalar potential to a metastable dS minimum.

In this paper, we revisit the question from the full ten-dimensional setup of Type IIB string theory, generalizing the analysis of
Maldacena-Nunez \cite{Maldacena:2000mw} by including extended localized sources in the gravity action. In particular we consider the traced-over Einstein equations, identifying the conditions for achieving de Sitter space in the non-compact dimensions for the cases
of fluxes, scalar fields and different localized sources, e.g. D-branes, anti D-branes and orientifold planes, in Type IIB with two-derivative gravity. We find that none of these ingredients satisfy the required condition, suggesting that one must consider additional terms in the gravity action.

One example of such additional terms is the set of higher-order curvature corrections. We perform an explicit calculation using an M-theory uplift, so as to simplify the form of the available fluxes. To study the effect of curvature corrections, we are forced to take an indirect route and instead consider a generalized correction to the action. We make an ansatz for the stress-energy tensor of the perturbative corrections, noting that the correction terms are built from curvatures. We explicitly find that positive curvature in the non-compact directions is only possible if curvature corrections are present and satisfy a certain inequality. 

We further find that the fluxes in any dS solution must be non self-dual, as is consistent with broken supersymmetry. These fluxes, combined with D-brane instantons, are enough to fix both the complex structure and Kahler moduli, including the volume modulus. In addition to this, the instantons are one possible source for the curvature correction terms required to give positive curvature to the non-compact space. We do not propose a specific form for these corrections, and as the complete set of supported corrections is not yet known, further conclusions cannot be made at this point. 

The structure of this paper is as follows: Sections 2 and 3 rederive the Gibbons-Maldacena-Nunez No-Go theorem, and apply it to bulk fields (fluxes and scalar fields) and localized sources.  In Section 4, we set up our M-theory calculation, which we perform in Section 5. We then examine the resulting equations of motion in Section 6 and 7, and discuss the origin of higher order curvature corrections in Section 8. We conclude our work with a short discussion of our results in Section 9.

\section{Einstein gravity in D dimensions}

Consider the following Einstein-Hilbert action coupled to matter in D spacetime dimensions:
\begin{equation}
\label{ActionE}
S_{\rm total}=\frac{1}{{\cal K}_D} \int d^{D}x \sqrt{-G_D}R_D+ \int d^Dx  {\cal L}_{\rm int},
\end{equation}
where ${\cal K}_{D}$ is the $D$-dimensional Newton constant, $R_D$ is the Ricci scalar in D dimensions, $G_D$ is the determinant of the D-dimensional metric $g_{MN}, M,N=0,..,D-1$, and ${\cal L}_{\rm int}$ is the Lagrangian for the local or global fields that couple to gravity. It can contain global fluxes, scalar fields, local sources and terms that describe graviton self coupling. In the Einstein equations,  ${\cal L}_{\rm int}$ enters through the stress-energy tensor 
\bg\label{TMNdef}
T_{MN}= -\frac{2}{\sqrt{-G_D}}\frac{\delta {\cal L}_{\rm int}}{\delta g^{MN}}. 
\nd
Variation of (\ref{ActionE}) with respect to $g^{MN}$ gives the following Einstein equation:
\bg\label{RMN}
R_{MN}=\frac{{\cal K}_{D}}{2} \left(T_{MN}-\frac{1}{D-2} g_{MN} T\right),
\nd
where $T$ is defined in the usual way, i.e.
\bg
T=g^{MN}T_{MN}.
\nd
Now we will split the geometry into two manifolds: $M_4$, spanned by coordinates $x^\mu, \mu=0,..,3$ and a 
transverse space ${\cal M}^{D-4}$, spanned by coordinates $x^m, m=4,..,D-1$. We want $M_4$ to describe our four dimensional 
non-compact space-time geometry and thus choose $(x^0,x^1,x^2,x^3)=(t,x,y,z)$, where $t$ is timelike. 
${\cal M}^{D-4}$ can be either a compact or non-compact $D-4$ dimensional manifold, 
described by spacelike coordinates $x^m$. We will often refer to $x^m$ and $x^\mu$ as describing internal and external directions respectively.  The line element is

\bg
ds_D^2=ds_4^2+ds_{D-4}^2\equiv g_{\mu\nu} dx^\mu dx^\nu+ g_{mn} dx^m dx^n .
\nd  
Now if the D-dimensional manifold has a direct product topology $M_4\times {\cal M}^{D-4}$,  then the  Ricci scalar for $M_4$ is:
\bg
R_4\equiv g^{\mu\nu} R_{\mu\nu}.
\nd
If $R_4>0$ we obtain a positive curvature spacetime, of which de Sitter space is one example, as is consistent with our universe. Alternatively, if  $R_4<0$, we have Anti-de Sitter type geometry, which is not consistent with the current universe.

 Taking the trace of (\ref{RMN}) in the $\mu,\nu$ directions, we get
 \bg\label{R4}
 R_4=-\frac{{\cal K}_D}{2(D-2)}\left[T^\mu_\mu (6-D)+4 T^m_m\right].
 \nd
  Thus for a positively curved spacetime, i.e. $R_4>0$, we must satisfy the condition:
 \bg\label{Condition1}
 \left(D-6\right)T^\mu_\mu>4T^m_m. 
 \nd
 Whatever the content of the Lagrangian, we must satisfy (\ref{Condition1}) if we are to obtain a positively curved four-dimensional universe. If we do not have a direct product space, but rather a warped product space, then the manifold cannot be nicely separated: $M_D\neq M_4\times {\cal M}^{D-4}$. However,  we can still try to obtain an effective four-dimensional space at low energies. In this case, the transverse dimensions are not accessible, which is possible if the size of ${\cal M}^{D-4}$ is small compared to the typical distance scale of interactions in $M_4$. We will separately address the case of a warped product space in the context of type IIB string theory in Section 3.2, where we will again see that the condition (\ref{Condition1}) plays a crucial role.
 
 We can now proceed to analyse different choices for the Lagrangian. 
 \subsection{Fluxes and scalar fields coupled to gravity}\label{flux}
We can reproduce the No-Go theorem of Gibbons \cite{Gibbons:1984kp, Gibbons:2003gb} and 
Maldacena-Nunez \cite{Maldacena:2000mw} by including fluxes in the Lagrangian. We consider the flux Lagrangian 
\bg
{\cal L}_{\rm int}^{F}=-\sqrt{-G_D} F_{a_1...a_q} F^{a_1...a_q},
\nd
where $F$ is a $q$-form. 
 The above Lagrangian leads to the following stress-energy tensor:
\bg
T_{MN}^{F}&=&-g_{MN}F^2+2 q F_{Ma_2..a_q}F^{a_2...a_q}_N.
\nd 
One can readily check that with the above form of the tensor, condition (\ref{Condition1}) will be satisfied if 
\bg \label{conditionF}
4(1-q)F^2>-F_{\mu a_2..a_q}F^{\mu a_2..a_q}q\left(D-2\right).
\nd
We will  consider two types of fluxes: the first type with legs only along the internal directions and the second type
with legs in $M_4$.   Also note that
 the overall minus sign in the Lagrangian is chosen to give 
positive energy,  i.e. $T_{00}>0$. For the first type of flux $a_i=m,n$ for all $i$ and
$F^2\ge 0$ with $F_{\mu a_2..a_q}F^{\mu a_2..a_q}=0$. Thus we find that condition (\ref{Condition1}) is 
{\it not} satisfied for $q>1$. 

If $q<4$ then all the legs will be along ${\cal M}^{D-4}$
since otherwise the isometries of $d=4$ Minkowski or de Sitter like space will be broken. Thus when we consider the second
type of flux which has legs in $M_4$, we will restrict to the case $q\ge 4$.  For $q\ge 4$ 
we will consider $4$ out of $q$ legs
along $M_4$ i.e. flux with legs in all the directions of $M_4$ and the rest of its legs along the internal directions. 
 With this condition on the fluxes, one obtains the following identities:
 \bg
  &&F^2=F_{a_1 a_2..a_q}F^{a_1 a_2..a_q}=C(q,4) F_{\mu_1..\mu_4 a_{5}..a_q}F^{\mu_1..\mu_4 a_{5}..a_q}\nonumber\\
  &&F_{\mu a_2..a_q}F^{\mu a_2..a_q}=C(q-1,3) F_{\mu_1..\mu_4 a_{5}..a_q}F^{\mu_1..\mu_4 a_{5}..a_q},\nd
where the coefficient $C(q, k)$ is defined by
 \bg\label{coefqk} C(q,k)\equiv \frac{q!}{(q-k)! k!}.
  \nd
 This in turn gives us 
  \bg
  F_{\mu a_2..a_q}F^{\mu a_2..a_q}=\frac{4}{q} F^2.
  \nd 
   Using the above relation and the fact that $F^2<0$, condition (\ref{Condition1}) will be satisfied if and only if
  \bg
  D<q+1.
  \nd
  Thus for $D>q+1$, we find that a $q$-form flux with legs in $M_4$ {\it does not} give rise to positive curvature for $M_4$.
   Any flux that preserves the desired isometries of $M_4$ can
   be written as a combination of the two types of fluxes described above. Thus, whatever the form of the flux, 
   $q$-form flux for $D>q+1$ {\it does not} give rise to positive curvature for $M_4$, as was first demonstrated by
   Maldacena and Nunez \cite{Maldacena:2000mw}. 
   
Next we consider scalar fields. The most general interaction Lagrangian for a scalar field interacting with gravity is given by 
\bg\label{scalar}
{\cal L}_{\rm int}^{\phi}=-\sqrt{-G_D} \left(\partial_{M}\phi\partial^M\phi+V(\phi)\right).
\nd
 Note that the overall minus sign is chosen so that when $V(\phi)=0$ (for example  massless fields with only kinetic energy), we get positive energy, i.e. $T_{00} > 0$. The stress-energy tensor is given by
 \bg
 T_{MN}^{\phi}=-g_{MN}\left(\partial_{K}\phi\partial^{K}\phi+V(\phi)\right)+2\partial_M\phi \partial_N\phi.
 \nd
  
 Then with the stress-energy tensor given above, the only way (\ref{Condition1}) is satisfied is if and only if
 \bg
 \partial_\mu\phi \partial^\mu\phi+V(\phi)>0 .
 \nd
 Now if we demand that the $M_4$ is isotropic in space but dependent on time, we readily find $\partial_\mu\phi
 \partial^\mu=g^{tt}\partial_t\phi \partial_t \phi<0$ since $g^{tt}<0$. 
 Thus {\it if} $V(\phi)<0$, $M_4$ {\it will not have positive curvature}. In type IIB string
 theory, which will be the focus of our study, the scalar axio-dilaton field $\tau$ has no potential and thus will not aid in
 constructing positive curvature.   
  
\subsection{Localized matter coupled to gravity}

Another possibility for the interaction Lagrangian is that of localized matter. For a p-dimensional object embedded in D-dimensional geometry, the most general Lagrangian that couples to the metric is the worldvolume Born-Infeld Lagrangian:
 
 \bg\label{mat}
 {\cal L}_{\rm int}^{\rm BI}= -T_p \sqrt{-\tilde{f}}\sqrt{g_{D-p-1}} \delta^{D-p-1}(x-\bar{x}),
 \nd    
 where $\tilde{f}$ is the determinant of the metric $\tilde{f}_{ab}$, defined in the following way:
\bg\label{detmet}
\tilde{f}_{ab}=f_{ab}+\widetilde{F}_{ab}, ~~~~~~ f_{ab}=g_{MN} \frac{\partial X^M}{\partial \sigma^a} 
 \frac{\partial X^N}{\partial \sigma^b}~~~~{\rm and}~~~~ \widetilde{F}_{ab}=F_{ab}+B_{ab}.\nd 
Here $T_p$ is the tension,  $F_{ab}$ is the worldvolume flux, $B_{ab}$ is the pullback of the background magnetic flux, $a,b=1,..,p+1$, and $\widetilde{F}_{ab}$  is  raised or lowered with the pullback
 metric $f_{ab}$. Also note that  $\delta^{D-p-1}(x-\bar{x})$ is the $(D-p-1)$-dimensional delta function, $x=\bar{x}$ is the location of the p-dimensional object, and $g_{D-p-1}$ is the determinant of the $(D-p-1)$-dimensional metric such that we have the normalization
 \bg
 \int d^{D-p-1}x \sqrt{g_{D-p-1}} \delta^{D-p-1}(x-\bar{x})=1.
 \nd
We have picked worldsheet parameters $\sigma^a=x^a, a=0,..,p-1$. $T_p$ can be considered as mass per unit length and thus it is typically positive. 
 
 If the Lagrangian is of the form (\ref{mat}) with positive mass term, i.e. $T_p>0$, one obtains:
 \bg\label{TMM}
&& T^{\enskip \mu\; (\rm BI)}_\mu =-T_p \frac{1}{\sqrt{-G_D}}\sqrt{-\tilde{f}} \sqrt{g_{D-p-1}}\tilde{f}_{ab}g^{\mu'\nu'}\frac{\delta \tilde{f}^{ab}}{\delta g^{\mu'\nu'}}\delta^{D-p-1}(x-\bar{x})<0\nonumber\\
&&  T^{\enskip m\; (\rm BI)}_m=-T_p \frac{1}{\sqrt{-G_D}}\sqrt{-\tilde{f}}\sqrt{g_{D-p-1}} \tilde{f}_{ab}g^{m'n'}\frac{\delta \tilde{f}^{ab}}{\delta g^{m'n'}}\delta^{D-p-1}(x-\bar{x})<0.
 \nd 
Using (\ref{TMM}) in (\ref{R4}) one readily sees that (\ref{Condition1}) is satisfied if $D<6$.  
 For $ D>6$ (\ref{Condition1}) is not automatically satisfied. In particular string theory gives $D = 10$ or 11 and thus we must have $T^m_m$ non-vanishing to obtain our four-dimensional positive curvature universe.
 
  String theory also allows negative tension objects, i.e. $T_p<0$, and higher-derivative terms in the low-energy effective action for gravity. Then, using the form of the  {\it localized} stress-energy tensor (\ref{TMM})  and adding the contributions from the fluxes, scalar fields and higher derivative terms, it may be possible to satisfy the condition (\ref{Condition1}). We will discuss this possibility in Sections 4 to 8.

\section{dS in Type IIB String Theory with Branes and Planes}

With a general understanding of gravitational coupling to fluxes and localized matter fields in $D$ dimensions, we will now consider the specific case of low-energy type IIB superstring theory with the following action in Einstein frame: 
\bg \label{Actions}
S_{\rm total}&=&S_{\rm SUGRA}+S_{\rm loc},
\nd
where
\bg\label{SUGRA}
S_{\rm SUGRA}&=&\frac{1}{2\kappa^2_{10}}\int d^{10}x \sqrt{- G_{10}}\left(R-\frac{\partial_M
\tau\partial^M\bar{\tau}}{2|{\rm
Im}\tau|^2}-\frac{|\hat{F}_5|^2}{4\cdot 5!}-\frac{G_3 \cdot \bar{G}_3}{12 {\rm Im}\tau}\right)\nonumber\\
&+&
\frac{1}{8i\kappa_{10}^2}\int \frac{C_4\wedge G_3\wedge \bar{G}_3}{{\rm Im} \tau}.
\nd 
Here $\tau=C_0+ie^{-\phi}$; $G_{10}={\rm det}~g_{MN}, M,N=0,..,9$; $g_{MN}$ is the metric in Einstein frame; $G_3=F_3-\tau H_3$; $F_3$ is the three-form RR flux, $H_3$ is the three-form NS-NS flux, and $\hat{F_5}$ is defined by
\bg
\hat{F_5} & = & F_5 - \frac{1}{2} C_2 \wedge H_3 + \frac{1}{2} B_2 \wedge F_3. 
\nd
 
 For the localized action we will consider Dp-branes and orientifold planes in various dimensions. The action for a Dp-brane is given by 
\bg \label{Action1}
S_{Dp}=-\int d^{p+1}\sigma \;T_p\;e^{\frac{\phi(p+1)}{4}}\;\sqrt{-\tilde{f}}+ 
\mu_p\int \left(C \wedge e^{\hat{F}}\right)_{p+1}.
\nd
Here $\tilde{f}$ is the same as in (\ref{mat}) and $C_{p+1}$ is the RR flux. As above,
$\widetilde{F}_{ab}$ is raised or lowered with the pullback metric $f_{ab}$. Note that the sign of $\mu_p$
determines whether we have a brane or an anti-brane. However both branes and anti-branes have positive tension $T_p>0$. 

On the other hand, for an orientifold, we have the action   

\bg \label{Action1s}
S_{Op}=-\int d^{p+1}\sigma \;T_{Op}e^{\frac{\phi(p+1)}{4}}\;\sqrt{-f}+\mu_{Op}\int C_{p+1} ,
\nd
where the orientifold has negative tension, i.e.  $T_{Op}<0$. Here $\mu_p$ is the charge of the Op-plane and  we have the  relation $|T_{Op}|=e^{-\phi}|\mu_{Op}|$. Also note that since the Op plane has negative charge, we have $\mu_p=e^{\phi}T_{Op}=-e^{\phi}|T_{Op}|$.

With the above localized action and the bulk supergravity action, we can write (\ref{Actions}) in the form (\ref{ActionE}) with the interaction Lagrangian being\footnote{The topological term cannot enter the stress-energy tensor since $\frac{\delta S_{\rm CS}}{\delta g^{MN}}=0$ where $S_{\rm CS}=\mu_p \int \left(C\wedge e^{\widetilde{F}}\right)_{p+1}$ is the Chern-Simons action. Therefore we omit it in the Lagrangians here. For Dp-branes $\hat{F}$ is not generally zero but Op-planes do not carry gauge fields, and have $\hat{F}$=0.} 
\bg
&&{\cal L}_{\rm int}={\cal L}_{\rm bulk}+{\cal L}_{\rm Dp}+{\cal L}_{\rm Op}\nonumber\\
&&{\cal L}_{\rm bulk}=\sqrt{-G_{10}}\left(-\frac{\partial_M
\tau\partial^M\bar{\tau}}{2|{\rm
Im}\tau|^2}-\frac{|\hat{F}_5|^2}{4\cdot 5!}-\frac{G_3 \cdot \bar{G}_3}{12 {\rm Im}\tau}\right)\nonumber\\
&& {\cal L}_{\rm Dp}= -T_p e^{\frac{\phi(p+1)}{4}}\sqrt{-\tilde{f}}\sqrt{g_{D-p-1}} \delta^{10-p-1}(x-\bar{x})\nonumber\\
&& {\cal L}_{\rm Op}= |T_{Op}| e^{\frac{\phi(p+1)}{4}} \sqrt{-f} \sqrt{g_{D-p-1}}\delta^{10-p-1}(x-\bar{x}).
 \nd
In the above ${\cal K}_{10}$ has been replaced by $2\kappa_{10}^2$. Using the above form of the Lagrangian we can readily obtain the stress-energy tensor (\ref{TMNdef}) and check whether the constraint (\ref{Condition1}) is satisfied or not.

To evaluate the trace of the stress-energy tensor, we will restrict the form of the fields to ensure Poincar\'e invariance in the non-compact spacetime. This way even without solving for the on-shell values of the fluxes and metric, we can check whether the inequality (\ref{Condition1}) is satisfied. These conditions are the following:
\vskip.1in \noindent
$\bullet$The fluxes $H_3$ and $F_3$ only have legs along  ${\cal M}^6$, and $\tau$ depends only on  $x^m$, the coordinates of ${\cal M}^6$. 
\vskip.1in \noindent
$ \bullet$ $\hat{F_5}$ will  have legs in the $x^\mu$ directions. Then by imposing self duality and  Poincar\'e invariance, one obtains the general form 
\bg \label{F5}
 \hat{F}_5=\left(1+\ast_{10}\right) d\alpha \wedge dt\wedge dx \wedge dy \wedge dz ,
 \nd 
 where $\alpha(x^M)$ is a scalar field which is a function of all coordinates $x^M,M=0,..,9$.  

Having laid down the required conditions, we will now analyze the individual cases with branes, anti-branes and orientifold planes. 

\subsection{Direct product space with Branes and Planes}
 We will first consider product spaces $M_{10}=M_4\times {\cal M}^6$ with branes and planes, where the transverse space $ {\cal M}^6$ can be either compact or non-compact. For $p=3$, we have  D3 or anti-D3 branes which  fill up $M_4$. Thus the induced metric is 
\bg
f_{ab}&=&g_{ab}, ~~ {\rm for}~~ a,b=\mu,\nu\nonumber\\
f_{ab}&=&0~~~~ ~{\rm for} ~~ a,b\neq \mu,\nu .
\nd 
Then we find \footnote{  Note that the upper indices here and elsewhere in this section have been raised with the metric $g^{MN}$, which is free of any warping in the case of a direct product space . For the warped compactifications studied in later sections, we will make the distinction between the warped metric and unwarped metric, where we introduce `tilded' quantities, $\tilde{A}^m$, that are defined with respect to the unwarped metric.}
\bg\label{Tmumu}
&&T^\mu_{\mu~(D3/\bar{D3})}= -T_3e^{\phi}\frac{\sqrt{-\tilde{f}}\sqrt{g_6}}{\sqrt{-G_{10}}}\left(4+\hat{F}^\mu_\mu\right)\delta^{6}(x-\bar{x})\nonumber\\
&&T^m_{m~(D3/\bar{D3})}=0.
\nd
However, since the flux $\hat{F}$ is anti-symmetric while the metric is symmetric, $\hat{F}^\mu_\mu=0$. Thus {\it neither} the D3 nor the anti-D3 brane tensor satisfies the constraint (\ref{Condition1}).

The results for D3 and anti-D3 branes can easily be generalized to Dp and anti-Dp branes with $p = 5,7$. For Poincar\'e invariance in the noncompact dimensions, we will fill up $M_4$ with the Dp or anti-Dp branes and the remaining worldvolume will fill up some $S^{p-3}$ cycle inside the transverse space ${\cal M}^{D-p-1}$. If $x^m,x^n$ denote coordinates of the cycle $S^{p-3}$, then we have 
\bg
f_{ab}&=&g_{ab}, ~~ {\rm for}~~ a,b=\mu,\nu, m,n\nonumber\\
f_{ab}&=&0~~~~ ~{\rm for} ~~ a,b\neq \mu,\nu, m,n.
\nd 
And we obtain 
 \bg\label{Tmumu}
 &&T^\mu_{\mu~(Dp/\bar{Dp})}=-T_{p}e^{\frac{\phi(p+1)}{4}}\frac{\sqrt{-\tilde{f}}\sqrt{g_{D-p-1}}}{\sqrt{-G_{10}}}\left(4+\hat{F}^\mu_\mu\right)\delta^{D-p-1}(x-\bar{x})\nonumber\\
 && T^m_{m~(Dp/\bar{Dp})}=-T_{p}e^{\frac{\phi(p+1)}{4}}\frac{\sqrt{-\tilde{f}}\sqrt{g_{D-p-1}}}{\sqrt{-G_{10}}}\left(p-3+\hat{F}^u_u\right)\delta^{D-p-1}(x-\bar{x}).
 \nd
Again, the worldvolume flux $\hat{F}$ is anti-symmetric while the metric is symmetric.  Hence $\hat{F}^\mu_\mu=0$. Using the form above, we can readily see that {\it neither} the Dp nor anti Dp-brane stress-energy tensor satisfies the constraint (\ref{Condition1}) for $p = 5,7$.  
 
Now for the five-form flux: using self-duality,  i.e. $|\hat{F}_5|^2=0$, one finds that the constraint (\ref{Condition1}) for the stress-energy tensor of the $\hat{F}_5$ will be satisfied if and only if 
\bg
\hat{F}_{\mu abcd}\hat{F}^{\mu abcd}>0.
\nd
However, using the form of the flux (\ref{F5}), it is straightforward to see that $\hat{F}_{\mu abcd}\hat{F}^{\mu abcd}<0$ 
 and thus the constraint (\ref{Condition1}) is not satisfied by the five-form flux. Alternatively, $\hat{F}_5$ can be
 written as a sum of two types of fluxes as described in section (\ref{flux}), and again we arrive at the same conclusion.
   
    Finally, using the condition that  $G_3$ has legs along ${\cal M}^6$ and $\tau$ only depends on $x^m$, 
    one finds that the stress-energy tensors for $G_3$ and $\tau$ do not satisfy the constraint (\ref{Condition1}). Since
    stress-energy tensors arising from fluxes, scalar fields or localized Dp or anti-Dp branes individually do not satisfy the constraint
    (\ref{Condition1}), the total stress-energy tensor for the entire system consisting of all these ingredients 
    will also not satisfy the constraint. 
    
    We can generalize the case for the localized Dp or anti-Dp branes to include smeared Dp or anti-Dp branes along
    the compact directions.\footnote{A discussion of smeared sources can be found in \cite{Blaback:2010sj, Caviezel:2009tu, Acharya:2006ne}. This procedure is a way to incorporate the global nature of charge cancellation into the 10d equations of motion, which are inherently local. Not all `smeared' solutions correspond to solutions of the full 10d equations.} The
    only difference in the smeared case is that the delta function in the stress-energy tensor (\ref{TMM}) will be replaced by
    some distribution i.e. $\delta(x-\bar{x})\rightarrow \Gamma(x^m)>0$. Smearing the branes in this fashion will allow one to compute the Ricci curvature {\it on the brane},
        which will be a finite quantity. Again, since $\Gamma(x^m)>0$, the stress-energy tensors will
    not obey the constraint  (\ref{Condition1}). In summary, we conclude that  local or non-local branes or anti-branes in the presence of global fields do not satisfy the condition (\ref{Condition1}) .     
    
     The only remaining case is the Op-planes. Orientifold planes are the loci of fixed points of some discrete symmetry group, arising from a $Z_2$ quotient of the theory combining worldsheet orientation reversal with an involution on the spacetime manifold \cite{Sen:1997gv}. The number of fixed points of 
 this orientifolding then gives the number of orientifold planes, which fill all the noncompact dimensions. 
 They have no gauge  fields on their worldvolume, and have negative fractional charge and tension. As the planes are fixed
 points of a symmetry group, their location in the internal space is fixed and cannot be arbitrarily chosen. Thus the planes
 are essentially localized and cannot be thought of as smeared objects.
 
 To construct an explicit gravity solution, we consider the localized action for the plane coupled with the bulk action.
 The tension of O3-planes taken to lie in $M_4$ is given by
 \bg \label{TmumuO3}
&& T^\mu_{\mu~(O3)}=4|T_{O3}|e^{\phi}\frac{\sqrt{-f}\sqrt{g_6}}{\sqrt{-G_{10}}}\delta^{6}(x-\bar{x})\nonumber\\
&& T^m_{m~(O3)}=0,
\nd
while for Op-planes with $p=5, 7$, assuming as above that the spacetime directions $M_4$ are filled, we find
 \bg\label{TmumuOp}
 && T^\mu_{\mu~(Op)}=4|T_{Op}|e^{\frac{\phi(p+1)}{4}}\frac{\sqrt{-f}\sqrt{g_{D-p-1}}}{\sqrt{-G_{10}}}  \delta^{D-p-1}(x-\bar{x})\nonumber\\
 && T^m_{m~(Op)}=|T_{Op}|e^{\frac{\phi(p+1)}{4}}\frac{\sqrt{-f}\sqrt{g_{D-p-1}}}{\sqrt{-G_{10}}}  \left(p-3\right)\delta^{D-p-1}(x-\bar{x}).
 \nd
Orientifolds have negative tension,  $T^\mu_{\mu~(Op)}>0$, 
so there is a possibility that the constraint (\ref{Condition1}) might be satisfied when O-planes are included. However we will see that this does not lead to positive curvature in four dimensions.  
 To see this
 first consider the Einstein equations arising from variation of the action (\ref{Actions}) with respect to the metric:
 
\bg \label{Ricci_Min}
&& R_{\mu\nu}=-g_{\mu\nu} \left[\frac{G_3 \cdot \bar{G_3}}{48\; {\rm
 Im}\tau}+\frac{\hat{F}_5^2}{8\cdot 5!}\right]+\frac{\hat{F}_{\mu
 abcd}\hat{F}_\nu^{\;abcd}}{4 \cdot 4!}
 +\kappa_{10}^2N_f \left(T_{\mu\nu}^{\rm loc}-\frac{1}{8} g_{\mu\nu} T^{\rm loc}\right),\nonumber\\
&& R_{mn}=-g_{mn} \left[\frac{G_3 \cdot \bar{G_3}}{48 \;{\rm
 Im}\tau}+\frac{\hat{F}_5^2}{8\cdot 5!}\right]+\frac{\hat{F}_{m
 abcd}\hat{F}_n^{\;abcd}}{4 \cdot 4!}+\frac{G_m^{\;bc}\bar{G}_{nbc}}{4\;{\rm Im}\tau}
 +\frac{\partial_m \tau \partial_n \tau}{2\;|{\rm Im}\tau|^2}\nonumber\\
 &&~~~~~~~~~~ + \kappa_{10}^2N_f \left(T_{mn}^{\rm loc}-\frac{1}{8} g_{mn} T^{\rm
 loc}\right),
 \nd
where $N_f$ is the number of localized objects contributing to $S_{\rm loc}$. Since we are considering manifolds which have the product form $M_{10}=M_4\times {\cal M}^6$, we have the following form for the metric: 

 \bg\label{metric}
 ds^2&=&
 g_{\mu\nu}(x^\mu) dx^\mu dx^\nu +g_{mn}(x^m) dx^m dx^n.
 \nd    
With this metric ansatz, taking the trace of the first equation in (\ref{Ricci_Min}) gives

\bg
 R_4(x^\mu)=- \frac{G_3 \cdot \bar{G_3}}{12\; {\rm
 Im}\tau}+\frac{\hat{F}_{\mu
 abcd}\hat{F}^{\mu abcd}}{4 \cdot 4!}
 +\frac{\kappa_{10}^2N_f}{2} \left(T_{\mu}^{\mu\; \rm loc}-T_{m}^{m\; \rm loc}\right).
\nd
The left-hand side is independent of $x^m$, and hence the right-hand side should be as well. It follows that we can evaluate the right-hand side at \emph{any} value of $x^m$, and so we are free to consider $x^m$ away from the localized Op-planes, where the local O-plane stress-energy tensor gives zero. As we have already studied, the flux and  local or smeared Dp or anti-Dp brane contributions to $R_4$ are negative definite. Thus we obtain 
\bg
R_{4}\le 0 .
\nd
 Since we have a product space $M_{10}=M_4\times {\cal M}^6$, $R_4$ is the Ricci scalar of $M_4$. Thus we conclude that neither Dp-branes, anti-Dp branes, nor Op-planes, in the presence of type IIB fluxes and scalar fields, give rise to positive curvature for $M_4$.

\subsection{Warped Product Manifold with Branes and Planes}
 Now we consider the  more general case where the ten-dimensional manifold is not a direct product space, but rather a warped product. We look for solutions to (\ref{Ricci_Min}) which take the following warped form:
 \bg\label{metricgen}
 ds^2&=&g_{\mu\nu}dx^\mu dx^\nu+g_{mn}dx^mdx^n\nonumber\\
 &=&e^{2A}\tilde{g}_{\mu\nu}dx^\mu dx^\nu +e^{-2A}\tilde{g}_{mn} dx^m dx^n,
 \nd    
 where $A(x^m)$ is a scalar function, $\tilde{g}_{\mu\nu}(x^\mu)$ is independent of internal coordinates $x^m$ while $\tilde{g}_{mn}(x^m)$ depends on $x^m$ . Now, using the ansatz (\ref{metricgen}) for the metric,
we get
  \bg\label{ricci_Min_a}
&& R_{\mu\nu}=\tilde{R}_{\mu\nu}-\tilde{g}_{\mu\nu}e^{4A}\widetilde{\triangledown}^2A ,
 \nd where the Laplacian is defined as
 \bg\label{Laplacian}
 \widetilde{\triangledown}^2=\widetilde{g}^{mn}\partial_m\partial_n
 +\partial_m\widetilde{g}^{mn}\partial_n+\frac{1}{2}\widetilde{g}^{mn}\widetilde{g}^{pq}\partial_n\widetilde{g}_{pq}
 \partial_m ,
 \nd
 and $\tilde{R}_{\mu\nu}$ is the Ricci tensor for the metric $\tilde{g}_{\mu\nu}$. Since the geometry is not a direct product, there is no notion of a separate four-dimensional space at all energies. If the internal space is compact and small, then at low energies we effectively have a four-dimensional non-compact space $\tilde{M}_4$ with metric $\tilde{g}_{\mu\nu}$. Then the condition  $\tilde{R}_4=\tilde{g}^{\mu\nu}\tilde{R}_{\mu\nu}>0$ states that $\tilde{M}_4$ has positive curvature. Thus, for a warped product geometry with metric of the form (\ref{metricgen}), we will restrict to the case where ${\cal M}^6$ is compact  and look for local and global fields in ten-dimensional type IIB theory that can give rise  to $\tilde{M}_4$ with positive curvature. 
 
 We take the trace of the first equation in (\ref{Ricci_Min}) and use the relation 
  (\ref{ricci_Min_a})
   to
  get
  \begin{eqnarray} \label{warp_eq_1a}
 \widetilde{\triangledown}^2e^{4A}&=&\tilde{R}_4+\frac{e^{2A}G_3 \cdot \bar{G_3}}{12\; {\rm
  Im}\tau}
 -\frac{e^{2A}\hat{F}_{\mu
  abcd}\hat{F}^{\mu abcd}}{4 \cdot 4!}+e^{-6A}\partial_me^{4A}\partial^me^{4A}\nonumber\\
 &+&\frac{\kappa_{10}^2}{2} e^{2A}\Big(\sum_i\left[T^m_{m~(Op/\bar{Op})i}-T^\mu_{\mu ~(Op/\bar{Op})i} \right]+\sum_j\left[T^m_{m~(Dp/\bar{Dp})j}-T^\mu_{\mu ~(Dp/\bar{Dp})j} \right]\Big) .\nonumber\\
 \end{eqnarray}
 Here $T^a_{a~(Op/\bar{Op})i}$ denotes the trace of the stress-energy tensor of the Op or anti-Op planes localized at $\bar{x}_i$, and similarly  $T^a_{a~(Dp/\bar{Dp})j}$ denotes the trace of the stress-energy tensor of the Dp or anti-Dp branes  at $\bar{y}_j$. The fluxes, branes, and planes, are related globally by charge cancellation, although we will not 
discuss the precise details here. We can integrate (\ref{warp_eq_1a}) over the compact internal manifold $\widetilde{\cal M}^6$ (which has the metric $\tilde{g}_{mn}$) to get
 
  \bg\label{miacons}
 C~ = &&~ \tilde{V}_6 \tilde{R}_4+\int d^6x\;\sqrt{\tilde {g_6}} {\cal I}_{\rm global}
 +\int d^6x\;\sqrt{\tilde {g_6}} \Big[ \frac{\kappa_{10}^2}{2} e^{2A}\Big(\sum_i\left[T^m_{m~(Op/\bar{Op})i}-T^\mu_{\mu ~(Op/\bar{Op})i} \right]\nonumber\\
 && +\sum_j\left[T^m_{m~(Dp/\bar{Dp})j}-T^\mu_{\mu ~(Dp/\bar{Dp})j} \right]\Big)\Big] , \nd
where $C=\int d^6x\;\sqrt{\tilde {g_6}}  \widetilde{\triangledown}^2e^{4A} $ is a constant and  we have defined ${\cal I}_{\rm global}$ and $\tilde{V}_6$ as
\bg\label{i3v6}
\nonumber  && {\cal I}_{\rm global}\equiv ~ \frac{e^{2A}G_3 \cdot \bar{G_3}}{12\; {\rm
    Im}\tau}
   -\frac{e^{2A}\hat{F}_{\mu
    abcd}\hat{F}^{\mu abcd}}{4 \cdot 4!}+e^{-6A}\partial_me^{4A}\partial^me^{4A}\ge 0, \\ 
    && \tilde{V}_6 \equiv ~ \int d^6x \sqrt{\tilde{g}_6}>0.
 \nd
 If ${\cal M}_6$ has no singularities {or the warp factor $e^{4A}$ is globally defined}, then $C=0$. However, in the presence of local sources classical gravity breaks down near the sources and this leads to physical singularities in the manifold.  To resolve these singularities, we can smear the Dp-anti-Dp branes while Op and anti-Op planes are by definition localized objects. If we remove the O planes entirely and only keep smeared branes, then   ${\cal M}_6$ will be regular and $C=0$. However as discussed in the previous section,  $T^m_{m~(Dp/\bar{Dp})}-T^\mu_{\mu ~(Dp/\bar{Dp})}\ge 0$,  and thus we get
     \bg
  \tilde{R}_4\le 0 .
  \nd
If we keep O planes, then there will be regions in the manifold with no classical gravity description. One can remove the singular points from the manifold leaving holes, but then $C\neq 0$\footnote{We thank Juan Maldacena for pointing this out. After the removal of points, $C$ becomes a boundary term. Additionally, removing the points means $T^m_{m~(Op/\bar{Op})i}-T^\mu_{\mu ~(Op/\bar{Op})i}=0$ but the effect of O planes is captured by the fluxes.}.  To obtain the exact value of $C$, one needs to know the metric near the singularity, but since classical gravity breaks down, we are unable to evaluate $C$. Thus, classical gravity is an incomplete description for a system containing O planes and we expect quantum corrections to resolve the classical singularity associated with the planes.       
 
 In summary,  {\it neither Dp nor anti-Dp branes with arbitrary worldvolume fluxes in the presence of type IIB fluxes and 
scalar fields result in positive curvature in four dimensions}. For direct product geometries, inclusion of Op or anti Op planes {\it also do not} give rise to positive curvature. For  {\it warped product geometries} arising in the  presence of Op or anti-Op planes,  classical two derivative gravity is insufficient and we must look for quantum corrections via higher-derivative gravity terms arising in string theory.

\section{Curvature Corrections and Background Solutions from M-theory}
\label{Mtysetup}

In the above sections we have argued that it is impossible to get a four-dimensional de Sitter spacetime in a 
ten-dimensional two-derivative gravity coupled to fluxes, scalar fields, D-branes and anti D-branes. 
With Orientifold-planes sourcing warped product manifolds, the classical gravity description is not sufficient 
to make a verdict one way or another. We need quantum corrections in the form of higher-curvature corrections to study 
the case with the Orientifold-planes. In fact string theory can have these corrections which, as we show below, could 
indeed help us to overcome the no-go theorem.

The analysis thus far has been done solely in the context of Type IIB string theory. However, the full set of quantum corrections in IIB is not known, and in addition there are many fields present which can complicate the analysis. To make the computations easier, we work in M-theory, where the bosonic field content is just the metric, $g_{MN}$, and the three-form, $C_{MNP}$, and make an ansatz for the form of the stress-energy tensor arising from any curvature corrections, given in (\ref{listofe}). A $T^2$ reduction of M-theory in the limit when the torus size goes to zero, will reproduce the answer for Type IIB theory.\footnote{Earlier studies using EOMs but without invoking quantum corrections may be found in \cite{noqotom}.}

We begin by setting up the M-theory uplift of the IIB system we are interested in. The action for M-theory is given by
\begin{equation}\label{Mtyaction}
S = S_{bulk}  + S_{brane} + S_{corr} ,
\end{equation}
where $S_{bulk} $ is the standard supergravity action for M-theory with a 3-form flux $C$ and corresponding field strength $G_4$, $S_{brane}$ is the contribution from $M2$-branes, and $S_{corr}$ is a curvature correction to the action. The supergravity and brane actions are given by
\begin{equation}
S_{bulk} = \frac{1}{2 \kappa^2} \int \mathrm{d}^{11}  x \;\sqrt{-g} \left[ R - \frac{1}{48} G^2\right] - \frac{1}{12 \kappa^2} \int C \wedge G \wedge G ,
\end{equation}
\begin{equation}
S_{brane} = - \frac{T_2}{2} \int \mathrm{d}^3 \sigma \sqrt{- \gamma} \left[ \gamma^{\mu \nu} \partial_\mu X^M\partial_\nu X^N g_{MN} -1 + \frac{1}{3!} \widetilde{\epsilon}^{\mu \nu \rho} \partial_\mu X^M\partial_\nu X^N \partial_\rho X^P C_{MNP}  \right] ,
\end{equation}
where $T_2$ is the tension of the $M2$-brane, $X^M$ denotes the worldsheet coordinates of the brane,  $\gamma^{\mu \nu}$ is the induced metric on the brane, and we have assumed a minimal coupling of the brane to the fluxes.

The corrections to the action are of the form $R^n$ or $G^n$ 
(or a combination thereof)\footnote{See for example \cite{deser} for more detail, up to four-point amplitudes, on this.}
and can come from several sources: instanton corrections, tree level $\alpha'$ corrections, and loop corrections.  
We delay a proper discussion of the $R^n$ terms to Section \ref{curvaturedisc}. To study the effect of these corrections, we first assume that $S_{corr}$ has two types of contributions: those that depend on the metric and are therefore non-topological, which we denote ${\hat S}_{ntop}$, and those that are topological and do not depend explicitly on the metric, ${\hat S}_{top}$.
In other words we have
\bg\label{socro}
S_{corr} ~ = ~ {\hat S}_{ntop} + {\hat S}_{top},\nd
where ${\hat S}_{top}$ can depend on the topological classes constructed out of the curvature form $R$. 
 
Both sets of corrections depend on the 
curvatures $R_{MNPQ}$ and $G_{MNPQ}$ of the metric $g_{MN}$ and the three-form field $C_{MNP}$ respectively, and we brand them curvature corrections. The contributions to ${\hat S}_{ntop}$ and ${\hat S}_{top}$ at lowest order in $\alpha'$  are known (see \cite{pisin} for example, as well as Section 8) and using these we can express ${\hat S}_{ntop}$ and ${\hat S}_{top}$ as
\bg\label{stopdan} 
&&{\hat S}_{top} ~=~ -T_2 \int C \wedge X_8 + {\cal S}_{top} (R, G)\nonumber\\
&& {\hat S}_{ntop} ~ =~ {T_2\over 9.2^{13}\cdot(2\pi)^4} \int d^{11} x \sqrt{-g} \left(J_0 - {1\over 2} E_8\right) + {\cal S}_{ntop}(R, G), \nd
where $X_8$ is the curvature correction  eight-form built completely with curvature two-form, such that $C \wedge X_8$ is a 
gravitational Chern-Simons term required to cancel the anomaly on the fivebrane worldvolume \cite{oai:arXiv.org:hep-th/9506126}; and $J_0$ and $E_8$ are given in \cite{pisin}. The 
additional contributions ${\cal S}_{ntop}$ and ${\cal S}_{top}$ are functions of both the curvatures ($R, G$). Some details of ${\cal S}_{ntop}$ and ${\cal S}_{top}$ have been worked 
out and they are given in \cite{deser} and \cite{jock} respectively. We will give a more complete discussion in Section 8.  
  
In Section \ref{EEqns}
we will make an ansatz for the variation of the correction terms with respect to the metric, which acts as an effective stress-energy tensor $T^{MN}_{corr}$, rather than deal with the action of the correction terms directly. In other words,  we will make an ansatz for
\bg\label{corfu}
T^{MN}_{corr} \equiv -{2\over \sqrt{-g}} {\delta S_{corr}\over \delta g_{MN}}\Big\vert_{g, C} ~=~ -{2\over \sqrt{-g}} {\delta {\hat S}_{ntop}\over \delta g_{MN}}\Big\vert_{g, C}~~, \nd 
where the subscript denotes a given choice of the metric and the three-form flux.

From the action \eqref{Mtyaction}, we obtain three key equations which govern the evolution of the system. The first is the Einstein equation,\footnote{We are assuming that the volume of the
internal fourfold is large so that an equation like \eqref{einsto} can be used to describe the metric there. This brings us to the issue of moduli stabilization, which will be discussed
towards the end of Section 7.} 
\begin{equation}\label{einsto}
R^{MN} - \frac{1}{2} g^{MN} R = T^{MN}  ,
\end{equation}
where $T^{MN}$ is the total stress-energy tensor coming from fluxes, brane sources and quantum or curvature corrections, and which we compute
in Section \ref{EEqns}. 
The second is the 
flux equation \cite{pisin},
\begin{equation}\label{fluxoo}
d \ast_{11} G = \frac{1}{2} G \wedge G + 2\kappa^2 \left( T_2 X_8 + \ast_{11} J\right) + S_G,
\end{equation}
where $J$ is the source term coming from $n_3$ M2-branes, $\ast_{11}$ is the Hodge star with respect to the warped metric unless mentioned otherwise,
and $S_G$ is the contribution from ${\cal S}_{ntop}$ and ${\cal S}_{top}$ in \eqref{stopdan} that we will discuss later.

The third equation is the $M2$-brane equation,
\begin{equation}\label{branulo}
\Box X^P + \gamma^{\mu \nu} \partial_\mu X^M \partial_\nu X^N \Gamma^{P_{MN}} = \frac{1}{3!} \epsilon^{\mu \nu \rho} \partial_\mu X^M \partial_\nu X^N  \partial_\rho X^Q {G^{P}} _{MNQ}  ,
\end{equation}
where $ \epsilon_{\mu \nu \rho} = \sqrt{- \gamma} \tilde\epsilon_{\mu \nu \rho}$. The source term at a spacetime position $x$ is related to the spacetime position $X$ of the 
brane, and is given by
\begin{equation}
J^{PQR}(x) = \frac{2 \kappa^2 n_3T_2}{ \sqrt{-g}} \int \mathrm{d}^3 \sigma \sqrt{- \gamma} \tilde\epsilon^{\mu \nu \rho} \partial_\mu X^P \partial_\nu X^Q \partial_\rho X^R \delta^{11} (x-X) .
\end{equation}
We would like to find a solution to these equations that is conformally de Sitter when brought to IIB, such that the IIB metric can schematically be written as
\begin{equation}
ds^2 = \frac{1}{t_{c} ^2} \eta_{\mu \nu} dx^\mu dx^\nu + ds^2 _{internal} ,
\end{equation}
where the time coordinate $t_c$ is conformal time, usually denoted $\tau$ or $\eta$, which in  the de Sitter space is related to physical time by
\begin{equation}
t_{c} \sim e^{ - t_{phys} } .
\end{equation}
It follows that the infinite future ($t_{phys} \rightarrow \infty$) is given by the limit $t_c \rightarrow 0$, as is the case during inflation. From this point onward we will drop the subscript $c$, and denote conformal time as $t$.

We make the following ansatz for the metric in M theory: 
\bg\label{mlift}
ds^2 ~& = & ~ {1\over (\Lambda(t) \sqrt{h})^{4/3}}(-dt^2 + \eta^{ij} dz_i dz_j) + h^{1/3}\left[{\tilde g_{mn} dy^m dy^n\over (\Lambda(t))^{1/3}} + (\Lambda(t))^{2/3} 
\vert dz\vert^2\right]\nonumber\\
& \equiv & ~ e^{2A(y, t)} (-dt^2 + \eta^{ij} dz_i dz_j) + e^{2B(y, t)} \tilde g_{mn} dy^m dy^n + e^{2C(y, t)} \vert dz\vert^2,\nd 
where $i, j = 1, 2$, $\tilde g_{mn}$ is the {\it unwarped} metric, $A, B$ and $C$ are warp factors that can be written in terms of  $\Lambda(t)$ and  $h(y^m)$, which we leave unspecified for the moment, and 
\bg\label{ltdz}
dz ~ \equiv ~ dx_3 + i dx_{11},\nd
so that the only time dependence in the system comes from $\Lambda(t)$. Specifically, the internal eight-dimensional manifold only depends on time via the warp factor $\Lambda(t)$ as we saw earlier, i.e.
\bg\label{amoktime}
ds^2_8 = {\tilde g_{mn}dy^mdy^n\over \Lambda^{1/3}(t)} + \Lambda^{2/3}(t) \vert dz\vert^2.\nd
This ansatz is chosen as the M-theory uplift for the solution we want to obtain in Type IIB, i.e. by shrinking the torus specified by coordinates ($z, \bar z$) or 
($x_3, x_{11}$) to zero size one may recover type IIB theory.  
It is a generalization of the ansatz considered in \cite{pisin}, and describes a system of M2-branes moving towards orbifold singularities  of the torus fibration of the fourfold (where the D7 fluxes are localized). This was developed as a first step towards an M theory uplift of D3/D7 \cite{dhhk}.


The IIB metric that follows from dimensional reduction of the M theory metric (\ref{mlift}) is given by
\bg\label{iibdsit2}
ds^2 = {1\over \Lambda(t) \sqrt{h}}(-dt^2 + \eta^{ij} dz_i dz_j + dx_3^2) + \sqrt{h}\tilde g_{mn} dy^m dy^n ,\nd
so that, taking $\Lambda(t) = \Lambda |t|^2$ (taking the absolute value to avoid any imaginary warping in the M-theory metric), we obtain
\bg
ds^2 = {1\over \Lambda t^2 \sqrt{h}}(-dt^2 + \eta^{ij} dz_i dz_j + dx_3^2) + \sqrt{h} \tilde g_{mn} dy^m dy^n . \nd
For this to be a dS solution, we demand that $\Lambda$  be strictly positive. We also require a suitably well-behaved functional form for $h(y)$, to avoid any pathology. However, for our purposes, we will leave its functional form to be completely general.

Turning now to the flux equations, the equation for the $G$-fluxes can be rewritten as:
\bg\label{gfeom}
&& D_M\left(G^{MPQR}\right) = {1\over \sqrt{-g}} \widetilde{\epsilon}^{PQRM_1....M_8}\left[{1\over 2\cdot (4!)^2}
G_{M_1....M_4}G_{M_5....M_8} + {2\kappa^2T_2\over 8!}(X_8)_{M_1....M_8}\right]\nonumber\\
&&~~~ + {2\kappa^2T_2n_3\over \sqrt{-g}}\int d^3\sigma \widetilde\epsilon^{\mu\nu\rho}\partial_\mu X^P\partial_\nu X^Q \partial_\rho X^R \delta^{11}(x-X) + 
{1\over \sqrt{-g}}\left({\delta {\cal S}_{ntop}\over \delta C_{PQR}} + {\delta {\cal S}_{top}\over \delta C_{PQR}}\right).\nonumber\\ \nd
The above equation is in general hard to deal with because of the quantum corrections etc. However the the $G$-fluxes are related to the membrane motion via the membrane EOM.  
In the limit where the membrane motion is very slow,  $\gamma_{\mu\nu}$, which is the pull-back metric, is simply equal to the spacetime metric given in \eqref{mlift}. This implies
\bg\label{implies}
G_{m\mu\nu\rho} ~ = ~\partial_m \left ( \frac{\widetilde\epsilon_{\mu \nu \rho}}{h \Lambda(t)^2}\right ), 
\nd
which shows that the spacetime part of the three-form field $C_{\mu\nu\rho}$ should be time-dependent to maintain a metric of the form \eqref{mlift} with a membrane fixed at a point on the eight-dimensional internal space. However to solve all the background equations we need more flux components. Let us then switch on the following three additional $G$-fluxes:
\bg\label{sogflux}
G_{mnpq} \equiv 4\partial_{[m}C_{npq]}, ~~~~ G_{mnpa} \equiv 3\partial_{[m}C_{npa]}, ~~~~ G_{mnab} \equiv 2\partial_{[m}C_{nab]} .\nd
To add some flexibility to the equations we seek to solve, and since we generically expect a mix of time-dependent and time-independent fluxes, we assume that the components $G_{mnpa}$ are time independent, whereas all other fluxes depend on the internal coordinates $y^m$, as well as on ($a, b)$ -- i.e. on ($x_3, x_{11}$) --  and the time $t$.  



\section{The Einstein Equations}
\label{EEqns}

In what follows we solve the Einstein equations \eqref{einsto} by including the general form of the stress-energy tensor $T_{MN}$ in Section \ref{EEqns}. This way we will be able to tabulate all the equations for the metric components satisfying \eqref{mlift}, in Section \ref{constraintssxn}. Subsequently, in Section \ref{fluxsxn}, we study the flux equations \eqref{fluxoo} and resulting consistency conditions.


\subsection{General Form of The Stress-Energy Tensor}
{
Like the action, the stress-energy tensor has 3 contributions:
\begin{equation}
T^{MN} = T^{MN} _{G} + T^{MN} _{corr} + T^{MN}_{B} ,
\end{equation} 
where $G$ is for G-flux, $corr$ is for correction, and $B$ is for brane.
As discussed in Section \ref{Mtysetup}, we will study the effect of higher-order curvature corrections to the action by making an ansatz for the resulting $T_{MN} ^{corr}$. Since our goal is to study solutions that are de Sitter in the non-compact dimensions, we are primarily concerned with tracking the time dependence of each component of the action and resulting Einstein equation. In line with this, we choose an ansatz for $T^{corr} _{MN}$ that allows us to keep track of the time dependence.}
The stress-energy contributions are then given by
\bg\label{listofe}
&& T^{MN}_{G} = \frac{1}{12} \left[ G^{MPQR} G^{N}_{PQR} - \frac{1}{8} g^{MN} G^{PQRS} G_{PQRS}\right]\\
&& T^{MN}_B (x) = -\frac{\kappa^2 T_2 n_3}{\sqrt{-g}} \int \mathrm{d}^3 \sigma \sqrt{- \gamma} \gamma^{\mu \nu} \partial_\mu X^{M} \partial_\nu X^{N} \delta^{11} (x- x_b)\\
&&T^{MN}_{corr} = \frac{-2}{ \sqrt{-g}} \frac{\delta {\hat S}_{ntop}}{\delta g_{MN}}\Big\vert_{g, C} \equiv  \displaystyle \sum_i [\Lambda(t)]^{\alpha_i + 1/3} \mathcal{C}^{MN,\,\, i}  ,
\nd
where again $x_b$ is the spacetime position of the brane (which is generically time dependent), and we have defined
\begin{equation}
\mathcal{C}^i _{MN} =  g_{MN} \mathcal{\tilde C}_i  - 2 \frac{\delta \mathcal{\tilde C}_i}{\delta g^{MN}} .
\end{equation}
 In the following sections we will attempt to search for solutions, by separately examining the $mn$, $ab$, and $\mu \nu$ components of the Einstein equation. Note that the scalars $\mathcal{\tilde C}_i$ are defined in terms of the \emph{unwarped metric}, such that the only dependence on warp factors in 
$\mathcal{C}_{MN}$ comes from the explicit factors of the warped metric $g_{MN}$.

\subsection{Internal ($m, n$) components}

We will start with the internal ($m, n$) components along the six-dimensional base. 
Two set of equations need to be solved now: the Einstein equation and the flux equation. For the Einstein equation we need the Einstein tensor from the M-theory metric 
\eqref{mlift}.
The Ricci tensor $R_{mn}$ is given by
\bg\label{ritti}
R_{mn}& =& \tilde R_{mn}+   3 \left[ 2\partial_{(m}A \partial_{n)}B - \partial_m A \partial_n A - \tilde g_{mn} \partial_k A \partial^k
B \right] + 4 \left[\partial_m B \partial_n B - \tilde g_{mn} \partial_k B \partial^k B
\right] \nonumber\\ 
&& - 3 D_{(m} \partial_{n)} A - 2 D_{(m}
\partial_{n)} C + 2 \left[ 2\partial_{(m}C \partial_{n)}B - \partial_m C \partial_n C
- \tilde g_{mn} \partial_k C \partial^k B \right] \nonumber\\
 && -4 D_{(m} \partial_{n)} B - \tilde g_{mn} \square B 
 + e^{2(B-A)} \left[ \ddot B + \dot A \dot B + 6 \dot B^2 + 2 \dot C \dot B \right] \tilde g_{mn} ,\nd
and the warped curvature scalar $R$ is given by
\bg\label{risca}
R & = & -e^{-2B} \left[ 10 
\square B + 6 \square A + 4 \square C + 20 \partial_m B \partial^m B
\right] - 3 e^{-2B} \left[ 4 \partial_m A \partial^m A + 8 \partial_m A \partial^m
B \right] \nonumber\\ && -2 e^{-2B} \left[ 3\partial_m C \partial^m C + 8
\partial_m B \partial^m C + 6 \partial_m A \partial^m C\right] + e^{-2B}~\tilde R\nonumber\\
&& + 2 e^{-2A}\left[6 \ddot B + 2 \ddot A + 2 \ddot C +
21 \dot B^2 + 6 \dot A \dot B + 12 \dot C \dot B + 2 \dot A \dot C
+ \dot A^2 + 3 \dot C^2\right] ,\nd
where remaining raising and lowering operations are done by the unwarped internal metric $\tilde g_{mn}$. The Einstein tensor $G_{mn}$ is found to be
\bg\label{gmnpo}
G_{mn} = \tilde G_{mn} - {\partial_m h \partial_n h \over 2h^2} + \tilde g_{mn} \left[{\partial_k h \partial^k h\over 4 h^2} - 6\Lambda h\right],\nd
where $\Lambda$ is the coefficient of $t^2$ in $\Lambda(t)$, and hence the above expression is independent of time.

To study the stress-energy tensor from the $G$-fluxes we have to first
express the various components of the $G$-fluxes $G_{MNPQ}$ in terms of their {\it unwarped} components $\widetilde{G}_{MNPQ}$ as:
\bg\label{unwarpy}
&& G^{012m} = \widetilde{G}^{012m} [\Lambda(t)]^{13/3} h^{5/3}, ~~~~~~~~~~~~~G^{012a} = \widetilde{G}^{012a} [\Lambda(t)]^{10/3} h^{5/3}\nonumber\\
&& G^{0mna} = \widetilde{G}^{0mna} [\Lambda(t)]^{4/3} h^{-1/3}, ~~~~~~~~~~~~G^{0mab} = \widetilde{G}^{0mab} [\Lambda(t)]^{1/3} h^{-1/3} \nonumber\\
&&  G^{mnpa} = \widetilde{G}^{mnpa} [\Lambda(t)]^{1/3} h^{-4/3}, ~~~~~~~~~~~~ G^{mnab} = \widetilde{G}^{mnab} [\Lambda(t)]^{-2/3} h^{-4/3} \nonumber\\ 
&& G^{0mnp} = \widetilde{G}^{0mnp} [\Lambda(t)]^{7/3} h^{-1/3}, ~~~~~~~~~~~~G^{mnpq} = \widetilde{G}^{mnpq} [\Lambda(t)]^{4/3} h^{-4/3}
\nd
where what we have done here is to simply isolate the warp factor dependences of $G^{MNPQ}$ and express its components in terms of $\widetilde{G}^{MNPQ}$. This also means that
$G_{MNPQ} \equiv \widetilde{G}_{MNPQ}$ by definition. We can also isolate the warp factor from the metric and write the determinant as
\bg\label{detdef} {\rm det}~g = - [\Lambda(t)]^{-14/3} h^{2/3} {\rm det}~\widetilde{g}.\nd 
The stress-energy tensor is easily expressed in the language of the unwarped $G$-fluxes \eqref{unwarpy} and the determinant \eqref{detdef}:
\bg\label{ggg}
{\cal T}_{mn}^{(G)} && =   \tilde g_{mn} {\partial_k h \partial^k h\over 4h^2} -  {\partial_m h \partial_n h\over 2h^2} + {1\over 4h}\left[\widetilde G_{mlka}\widetilde G_n^{~~lka} - {1\over 6} \tilde g_{mn} \widetilde G_{pkla} \widetilde G^{pkla}\right]\\
&& ~~ + {\Lambda(t)\over 12 h}\left[\widetilde G_{mlkr} \widetilde G_n^{~~lkr} - {1\over 8}\tilde g_{mn} \widetilde G_{pklr} \widetilde G^{pklr}\right] + {1\over 4h \Lambda(t)}
\left[\widetilde G_{mlab} \widetilde G_n^{~~lab} - {1\over 4}\tilde g_{mn} \widetilde G_{pkab} \widetilde G^{pkab}\right].\nonumber
\nd
The stress-energy tensor from the membrane (M2 brane) will not contribute however. This is because the stress-energy tensor, given by \cite{pisin},
\bg\label{emtmem}
{\cal T}_{mn}^{(B)} = -\kappa^2 T_2 n_3 \tilde g_{pm} \tilde g_{qn} \frac{h^{1/3}[\Lambda(t)]^{5/3}}{\sqrt{\tilde g}} \int d^3\sigma \sqrt{-\gamma}\gamma^{\mu\nu}\partial_\mu X^p \partial_\nu X^q \delta^{11}(x - X),
\nd
where $\tilde g$ is the determinant of the metric in the $m,n$ directions, vanishes in the limit where the membrane motion is very slow. The only other contribution will be from the correction terms, which, using $g_{mn} = e^{2B} \tilde g_{mn}$, gives
\bg\label{correctedTmn}
{\cal T}_{mn}^{corr} & = &    h^{1/3} \displaystyle \sum_i [\Lambda(t)]^{\alpha_i }  \mathcal{\tilde C}^i _{mn}.
\nd
The equation that we need to solve now is 
\bg\label{einsto}
G_{mn} =  {\cal T}_{mn}^{(G)} + {\cal T}_{mn}^{corr} .\nd
This can be split into a time-independent piece,
\bg \label{TImncorrected}
\tilde G_{mn} - \tilde g_{mn}  6\Lambda h=  \frac{1}{4h} \left [ \widetilde G_{mlka} \widetilde G_n^{\enskip lka} - \frac{1}{6} \tilde g_{mn} \widetilde G_{pkla} \widetilde G^{pkla} \right ] +   h^{1/3} \displaystyle \sum_ {\alpha_i =0} \mathcal{\tilde C}^i _{mn},\nd
where we made use of our assumption that the $G_{mnpa}$ are time independent, and a time-dependent piece given by
\bg\label{temcorrected}
&& \frac{\Lambda(t) }{12h} \left [ \widetilde G_{mpqr} \widetilde G_n^{ \enskip pqr} - \frac{1}{8} \tilde g_{mn} \widetilde G_{pqrs} \widetilde G^{pqrs} \right] + \frac{1}{4 h \Lambda(t)} \left [ \widetilde G_{mpab} \widetilde G_n^{\enskip pab} - \frac{1}{4} \tilde g_{mn} \widetilde G_{pqab} \widetilde G^{pqab} \right] \nonumber\\  
&& ~~~~~~~  +     h^{1/3} \displaystyle \sum_{\alpha_i \neq 0} [\Lambda(t)]^{\alpha_i } \mathcal{\tilde C}^i _{mn}~ = ~ 0.\nd
Note that at this stage the only possible way $G_{mnpr}$ and $G_{mnab}$ can also be time independent and yet still satisfy \eqref{temcorrected} is if the $\alpha_i$ are allowed to take
the values
\bg\label{alpval}
\alpha_i ~ = ~ (1, -1, 0, 0, ....0) .\nd
It is not clear we can have this condition for our case, and so we will assume that the only time-independent components of the $G$-fluxes are $G_{mnpa}$. 

\subsection{Internal ($a, b$) components}
The Ricci tensor for the ($a, b$), i.e. the $x^3$ and $x^{11}$ components, is given by

\bg\label{rikk} 
R_{ab} = && -\delta_{ab}e^{2(C-B)}\left[ \square C + 3 \partial_m C
\partial ^m A + 4 \partial_m C \partial^m B + 2 \partial_m C \partial^m C \right] \nonumber\\
&& ~~~~~~~ + \delta_{ab} e^{2(C - A)} \left[ \ddot C + \dot A \dot C +
6 \dot C \dot B + 2 \dot C^2 \right] , \nd
which can be used to compute the Einstein tensor $G_{ab}$. For the M-theory metric \eqref{mlift}, $G_{ab}$ is given by
\bg\label{gab}
G_{ab} ~ = ~ \delta_{ab} \Lambda(t) \left[-{\tilde R \over 2} - 9h \Lambda + {\tilde g^{pk} \partial_p h \partial_k h\over 4h^2}\right] , \nd
where we note that there is an overall time dependence given by $\Lambda(t)$. The stress-energy tensor due to the fluxes is given by
\bg\label{gggg} 
\nonumber  {\cal T}_{ab}^{(G)} &&=  {\Lambda(t)\over 12 h}\left[\widetilde G_{amnp}\widetilde G_b^{~~mnp} - \delta_{ab} {\widetilde G_{mnpc} \widetilde G^{mnpc}\over 2}
+ \delta_{ab} {3{\tilde{g}}^{mp}\partial_m h \partial_p h\over h} \right] \\&&+ {1\over 4h}\left[\widetilde G_{acmn}\widetilde G_b^{~~cmn} - {1\over 4}\delta_{ab}\widetilde G_{mncd}\widetilde G^{mncd}\right]  - \delta_{ab}{[\Lambda(t)]^2 \over 4\cdot 4! h}\widetilde G_{mnpq}\widetilde G^{mnpq}.
\nd
The interesting thing about the above formula is that the time dependence of the first term (involving $\widetilde{G}_{mnpa}$) 
is exactly  the same as the time dependence of the 
$G_{ab}$. This means that the $\widetilde{G}_{mnpa}$ components can remain time independent, as we had earlier.  The correction term contribution to the stress-energy tensor for the $(a, b)$ directions is
\bg \label{khattacorr}{\cal T}_{ab}^{corr}  =     h^{1/3} \displaystyle \sum_i [\Lambda(t)]^{\alpha_i+1 } \mathcal{\tilde C}^i _{ab} . \nd
As before, we can write the resulting Einstein equation as a time-independent expression (where we collect the terms linear in $\Lambda(t)$):
\bg\label{qottocorr}
&& \left (\frac{\tilde R}{2} + 9 h \Lambda \right) \delta_{ab} + \frac{1}{12h} \left [\widetilde G_{amnp} \widetilde G_b^{\enskip mnp} - \delta_{ab}\frac{ \widetilde{G}_{mnpc} 
\widetilde{G}^{mnpc}}{2} \right] \nonumber\\ 
&& ~~~~~~~~~~~~~+   h^{1/3} \displaystyle \sum_{\alpha_i =0} \mathcal{\tilde C}^i _{ab} ~ = ~ 0,
\nd
and a time-dependent expression:
\bg\label{tem2corr}
 && \frac{1}{4h} \left [\widetilde G_{acmn} \widetilde G_b^{\enskip cmn} - \frac{1}{4} \delta_{ab} \widetilde G_{mncd} \widetilde G^{mncd} \right ] 
- \delta_{ab} \frac{[\Lambda(t)]^2}{4 \cdot 4! h} \tilde G_{mnpq} \tilde G^{mnpq}\nonumber\\ 
&& ~~~~~~~~~ + h^{1/3}  \sum_{\alpha_i \neq 0} \left [\Lambda(t) \right ]^{\alpha_i +1} \mathcal{\tilde C}^i _{ab} ~ = ~ 0.
\nd
Once again, we must assume $G_{mnpq}$ and $G_{mnab}$ are time dependent in such a way as to solve \eqref{tem2corr}. 
Thus the conclusion of this section is perfectly consistent with the 
conclusions of the previous section. 

\subsection{Spacetime ($t, z_1, z_2$) components}
%

We now  study the spacetime components. The curvature tensors $R_{00}$ and $R_{ij}$ are given by
\bg\label{rmumu}
R_{ij} &=& -\eta_{ij} e^{2A-2B}\left[\square A + 3\partial_m A \partial^m A + 4 \partial_mA \partial^m B + 2 \partial_m A \partial^m C\right] \\
\nonumber && + \left(\ddot A + 6 \dot A \dot B + {\dot A}^2 + 2 \dot A \dot C\right)\eta_{ij} \\
R_{00} &= &e^{2A-2B}\left[\square A + 3\partial_m A \partial^m A + 4 \partial_mA \partial^m B + 2 \partial_m A \partial^m C\right] \\
 &&- \left[2\ddot A + 6(\ddot B + {\dot B}^2 - \dot A \dot B) + 2(\ddot C + {\dot C}^2 - \dot A \dot C)\right] ,
 \nd
using which the Einstein tensor $G_{\mu\nu}$ is found to be
\bg\label{gmunu}
G_{\mu\nu} ~ = ~ -{\eta_{\mu\nu}\over \Lambda(t)} \left[{\tilde R\over 2h} + {\tilde g^{mk} \partial_k h \partial_m h\over 4h^3} - {\square h \over 2h^2} + 3\Lambda\right] , \nd
where we see that the overall time dependence is provided by $1/\Lambda(t)$. 
The above equation should be balanced by the stress-energy tensor from the $G$-flux and corrections, as well as from the membrane. The latter term is there because the almost static membrane \emph{does} contribute to the stress-energy tensor along the spacetime directions.

The stress-energy tensor from the $G$-flux is given by
\begin{eqnarray}\label{enermog}
{\cal T}_{\mu\nu}^{(G)} &=& -\eta_{\mu\nu} \left[\frac{(\partial h)^2}{4  \Lambda(t) h^3} +\frac{\widetilde G_{mnpa}\widetilde G^{mnpa}}{4! \Lambda(t) h^2}   
+ {\widetilde G_{mnpq}\widetilde G^{mnpq}\over 4\cdot 4! h^2} 
  + {\widetilde G_{mnab}\widetilde G^{mnab}\over 16 h^2 [\Lambda(t)]^2} \right]. 
 \end{eqnarray}
As expected, ${\cal T}_{\mu\nu}^{(G)}$ has a piece that scales as $1/\Lambda(t)$, so we should be able to maintain the time independence of the $G_{mnpa}$ components. 

The stress-energy tensor coming from the correction terms can be found to be
\bg\label{qotom3}
T_{\mu \nu}^{corr}  =    h^{-2/3} \displaystyle \sum_i [\Lambda(t)]^{\alpha_i-1 } {\mathcal{\tilde C}^i}_{\mu \nu} .
\nd
Finally we will need the stress-energy tensor for the static membrane. The EOM of the worldvolume metric gives us, in the case where the brane is moving very slowly,
\bg\label{hinsro} 
\gamma_{\mu\nu} ~ = ~ \partial_\mu X^M \partial_\nu X^N g_{MN} ~ \approx g_{\mu \nu} ~= ~ {\eta_{\mu\nu}\over [\Lambda(t) \sqrt{h}]^{4/3}}.\nd
Using this we can show that the stress-energy tensor is given by
\bg\label{khatta} 
{\cal T}_{\mu\nu}^{(B)} = -{ \kappa^2 T_2 n_3 \over h^2 \Lambda(t) \sqrt{\tilde{g}}}\delta^8(x-X) \eta_{\mu \nu} , \nd
which is again suppressed by $1/\Lambda(t)$, confirming the time independence of the components $G_{mnpa}$. 

Again, we can split the full Einstein equation into a time-independent part:
\bg\label{juasha} 
\left( {\tilde{R}\over 2h} - {\square h \over 2h^2} + 3\Lambda \right) =  {\widetilde{G}_{mnpa}\widetilde{G}^{mnpa}\over 4!h^2} + { \kappa^2 T_2 n_3 \over h^2  \sqrt{\tilde{g}}}\delta^8(x-X) -  \frac{1}{ 3 h^{2/3}} \sum_{\{\alpha_i\}= 0} {\mathcal{\tilde C}}_{\mu} ^{\mu, i}  \nd
where we have traced over the $\mu,\nu$ components using $\eta_{\mu \nu}$, and a time-dependent part:
\bg\label{jolashor}
\eta_{\mu \nu} \left[ {\widetilde{G}_{mnpq}\widetilde{G}^{mnpq}\over 4\cdot 4! h^2} + {\widetilde{G}_{mnab}\widetilde{G}^{mnab}\over 4!h^2 \Lambda(t)^2} \right] - 
\frac{1}{h^{2/3}} \sum_{\{\alpha_i\} \ne 0}[\Lambda(t)]^{\alpha_i - 1} {\mathcal{\tilde C}^{i}} _{\mu \nu} =0.\nd

\section{Analysis of the EOMs and Consistency Conditions}
\label{constraintssxn}

We have now split the Einstein equations into 6 equations, 3 of which are time dependent, and 3 of which are time independent. To deduce the properties of these equations, it suffices to look at the traced over form of each. The traced-over time independent equation for the spacetime ($\mu, \nu$) components is
\bg\label{firstconst} 
\left( {\tilde{R}\over 2h} - {\square h \over 2h^2} + 3\Lambda \right) =  {\widetilde G_{mnpa}\widetilde G^{mnpa}\over 4!h^2} + \frac{\kappa^2 n_3T_2 \delta^8 (x - X)}{h^2 \sqrt{\tilde g}} - \frac{1}{3 h^{2/3}} \sum_{\{\alpha_i\}= 0} {\mathcal{\tilde C}}_{\mu} ^{\mu,\,\, i} , \nd
whereas for the internal ($m, n$) components, it is
\bg\label{secondconst}36h \Lambda  +  h^{1/3} \sum_{\{\alpha_i\}=0} \mathcal {\tilde C}_{m} ^{m, \,\, i}  = \tilde{G}_{m} ^{m} . \nd
Note that the flux contribution in \eqref{TImncorrected} is traceless, so it doesn't appear in the above equation. Finally, for the internal ($a, b$) components the trace equation is
\bg\label{thirdconst} {\tilde{R}\over 2}  +  9h \Lambda  + \frac{ h^{1/3}}{2} \sum_{\{\alpha_i\}=0} \mathcal{\tildeC}_{a}^{a, \,\, i} =0 , 
\nd
where again the flux contributions from \eqref{qottocorr} do not enter. The last two equations, \eqref{secondconst} and \eqref{thirdconst}, are quite similar and
can be rewritten as
\begin{eqnarray}\label{jom}
\sum_{\{\alpha_i\}=0} \mathcal{\tildeC}_{m}^{m, \,\, i}& = & - \frac{2}{h^{1/3}}( \tilde R + 18 h \Lambda), \\
\sum_{\{\alpha_i\}=0} \mathcal{\tildeC}_{a}^{a, \,\, i} & = & - \frac{1}{h^{1/3}} (\tilde R + 18 h \Lambda),
\end{eqnarray}
from which we can read off that
\begin{eqnarray}
\sum_{\{\alpha_i\}=0} \mathcal{\tildeC}_{m}^{m, \,\, i}& = & 2 \sum_{\{\alpha_i\}=0} \mathcal{\tildeC}_{a}^{a, \,\, i} .
\end{eqnarray}
Using (\ref{secondconst}) and (\ref{thirdconst}) we can also write
\begin{eqnarray}
\tilde R & = & - 18 h \Lambda - h^{1/3} \left (\frac{1}{2} \sum_{\{\alpha_i\}=0} \mathcal{\tildeC}_{a}^{a, \,\, i} + \frac{1}{4} \sum_{\{\alpha_i\}=0} \mathcal{\tildeC}_{m}^{m, \,\, i} 
\right), \end{eqnarray}
which allows us to rewrite the constraint (\ref{firstconst}) as
\begin{eqnarray}\label{conoklo} 
- \Box h & = & \frac{\widetilde G_{mnpa} \widetilde G^{mnpa}}{12} + 12 h^2 \Lambda + \frac{2 \kappa^2 n_3 T_2 \delta^8(x - X)}{\sqrt{\tilde g}}\nonumber\\
&& + h^{4/3} \left (\frac{1}{2} \sum_{\{\alpha_i\}=0} \mathcal{\tildeC}_{a}^{\enskip a, \,\, i}  + \frac{1}{4} \sum_{\{\alpha_i\}=0} \mathcal{\tildeC}_{m}^{\enskip m, \,\, i} - \frac{2}{3}\sum_{\{\alpha_i\}= 0} {\mathcal{\tilde C}}_{\mu} ^{\enskip \mu,\,\, i}  \right).
\end{eqnarray}
There are three further equations that arise from \eqref{qottocorr} in the limit when $a \ne b$, $a = b = 3$ and $a = b = 11$ respectively. These are 
\bg\label{threeeom}
&& \widetilde{G}_{amnp}\widetilde{G}_b^{mnp} + 12 h^{4/3} \displaystyle \sum_{\{\alpha_i \}=0} \mathcal{\tilde C}^i _{ab} ~ = ~ 0 , \nonumber\\
&& \widetilde{G}_{3mnp}\widetilde{G}_3^{mnp} - \widetilde{G}_{11,mnp}\widetilde{G}_{11}^{mnp} = 
24 h^{4/3} \sum_{\{\alpha_i\}=0} \left({1\over 2} \mathcal{\tildeC}_{a}^{a, \,\, i} - \mathcal{\tildeC}_{33}^{i}\right) , \nonumber\\
&& \widetilde{G}_{3mnp}\widetilde{G}_3^{mnp} - \widetilde{G}_{11,mnp}\widetilde{G}_{11}^{mnp} = - 24 h^{4/3} \sum_{\{\alpha_i\}=0} \left({1\over 2} \mathcal{\tildeC}_{a}^{a, \,\, i}
- \mathcal{\tildeC}_{11,11}^{i}\right). \nd
If we now consider integrating equation \eqref{conoklo}  over the compact eight-dimensional manifold, we see that the LHS integrates to zero as the warp factor $h$ is a globally defined quantity, and we get 
\bg\label{conokl} 
&& {1\over 12}  \int d^8 x \sqrt{\tilde g}~ \widetilde{G}_{mnpa}\widetilde{G}^{mnpa}  + 12 \Lambda \int d^8 x \sqrt{\tilde g}~ h^2  + 2 \kappa^2 T_2 n_3\nonumber\\
&& ~ + \int d^8x \sqrt{\tilde g} h^{4/3} \left (\frac{1}{2} \sum_{\{\alpha_i\}=0} \mathcal{\tildeC}_{a}^{a, \,\, i}  + \frac{1}{4} \sum_{\{\alpha_i\}=0} \mathcal{\tildeC}_{m}^{m, \,\, i} - \frac{2}{3}\sum_{\{\alpha_i\}= 0} {\mathcal{\tilde C}}_{\mu} ^{\mu,\,\, i}  \right)  = 0.\nd  
In the absence of fluxes and higher-curvature corrections the above equation implies that 
the simplest solution will be $\Lambda = 0$, i.e. a four-dimensional Minkowski space. This conclusion {\it cannot} be changed by the insertions of the type IIB Orientifold-planes precisely because they become smooth geometries\footnote{The ``twisted sector'' states appear precisely from smoothing the geometry in M-theory. The higher curvature 
term $C \wedge X_8$ provides the gravitational couplings on the corresponding type IIB Orientifold-planes as will be briefly discussed above \eqref{d7o7}. The rest of the 
curvature terms from ${\cal S}_{top}$ and $\hat{S}_{ntop}$ in \eqref{stopdan} contribute to the higher curvature terms on the Orientifold-planes beyond the Chern-Simons terms 
of \eqref{d7o7} and \eqref{m2b}. For more details see \cite{mukdas1, mukdas2}.} 
in M-theory and therefore {\it cannot} change the sign of $\Lambda$ in the absence of any corrections. 
In the presence of fluxes, and in the presence or
absence of the higher-curvature corrections, it is not difficult to see that the $\Lambda < 0$ solution is favored. However to allow a $\Lambda > 0$ solution from \eqref{conokl}, it is 
{\it at least} necessary to have the higher curvature corrections, because 
the first three terms in \eqref{conokl} are positive definite.
Moreover, if all the curvature corrections in \eqref{conokl} add up to some positive value, a $\Lambda > 0$ solution will 
again be impossible.

This means that for a $\Lambda > 0$ solution to exist, the curvature terms in \eqref{conokl} should integrate to a negative definite value. This conclusion should be valid for all possible
choices of the globally-defined warp factor $h$ and the internal metric $\tilde g_{mn}$. In particular, 
for certain choices of the warp factor the fluxes may be localized over a small patch on the internal manifold (for 
example like the type IIB seven-brane solution). Then the integral condition on 
the higher-curvature terms will have to be realized at every such patch on the internal manifold. 
On a small patch, since there is no local transformation that can make the metric flat everywhere,  ${\mathcal{\tilde C}}_{M} ^{M,\,\, i}$ can be viewed as the 
expectation or the average value on the patch, or more explicitly:
\bg\label{average}
\langle{\mathcal{\tilde C}}_{M} ^{M,\,\, i}\rangle \equiv \int d^8x \sqrt{\tilde g}~ h^{4/3} {\mathcal{\tilde C}}_{M} ^{M,\,\, i} . \nd
In other words, for a solution to exist we must have the following condition 
 \begin{eqnarray}
 \label{explicitconst} \frac{1}{2} \sum_{\{\alpha_i\}=0} \langle\mathcal{\tildeC}_{a}^{a, \,\, i}\rangle  
+ \frac{1}{4} \sum_{\{\alpha_i\}=0} \langle\mathcal{\tilde C}_{m}^{m, \,\, i}\rangle 
- \frac{2}{3}\sum_{\{\alpha_i\}= 0} \langle{\mathcal{\tilde C}}_{\mu} ^{\mu,\,\, i}\rangle & < & 0.
 \end{eqnarray}
Since $T_{mn}^{corr} \sim \tilde C^i_{mn}$, this equation is almost analogous to (\ref{Condition1}) but expressed in the language of curvature corrections.\footnote{One subtlety however is that this constraint arises from the Einstein equations of an 11-dimensional M theory, in which $\mu$ runs from 0 to 2, while in (\ref{Condition1}) it runs from 0 to 3, so the numerical factors are not expected to be the same in both expressions. We would have to redo the calculation in IIB to get the same expression. However
in both cases the condition is that the four-dimensional curvature upon compactification be positive. 
} This makes 
sense because only these corrections will allow us to overcome the Gibbons-Maldacena-Nunez \cite{Gibbons:1984kp, Gibbons:2003gb, Maldacena:2000mw} no-go theorem. 
{Under this assumption, \eqref{explicitconst} gives non-trivial constraints on the curvature corrections required to have a four-dimensional de Sitter solution in Type IIB theory.}

The curvature terms may be further constrained if we look at the time-dependent equations. These equations are
\begin{eqnarray}
&&  {\widetilde{G}_{mnpq}\widetilde{G}^{mnpq}\over 4} + {\widetilde{G}_{mnab}\widetilde{G}^{mnab}\over \Lambda(t)^2} = 
{8}{h^{4/3}}\sum_{\{\alpha_i\} \ne 0}[\Lambda(t)]^{\alpha_i - 1} {\mathcal{\tilde C}} _{\mu} ^{\mu,\,\, i} ,\\
&&  \widetilde{G}_{acmn} \widetilde{G}^{acmn}  - \frac{[\Lambda(t)]^2}{6} \widetilde{G}_{mnpq} \widetilde{G}^{mnpq}  =   - 8h^{4/3}  \sum_{\alpha_i \neq 0} \left [\Lambda(t) \right ]^{\alpha_i +1} {\mathcal{\tilde C}} _{a} ^{a, \,\, i },\\
\label{temcorrected} 
&& \frac{\Lambda(t) }{6}  \widetilde{G}_{pqrs} \widetilde{G}^{pqrs}  - \frac{1}{ \Lambda(t)} \widetilde{G}_{mpab} \widetilde{G}^{mpab}   =   - 8h^{4/3} \displaystyle \sum_{\alpha_i \neq 0} [\Lambda(t)]^{\alpha_i } {\mathcal{\tilde C}} _{m} ^{m, \,\, i} .
 \end{eqnarray}
From the first equation above, and noting that both the terms on the LHS are positive definite, we deduce one new condition on the corrections by integrating over the 
eight-dimensional manifold:
\begin{equation}\label{corrcorr1}
 \sum_{\{\alpha_i\} \ne 0} a^{\alpha_i} \langle{\mathcal{\tilde C}} _{\mu} ^{\mu, \,\, i}\rangle ~ >  ~ 0 ,
\end{equation}
where $a \equiv \Lambda(t_a)$ for a fixed $t_a$. In fact \eqref{corrcorr1} will be an infinite set of constraints because, {due to its time dependence,}  $a^{\alpha_i}$ can take any (positive) values including arbitrary 
fractional numbers.
Note that 
\bg\label{simpol}
\langle{\mathcal{\tilde C}} _{\mu} ^{\mu, \,\, i}\rangle >0
\nd
will always solve \eqref{corrcorr1} if the $\alpha_i$ appearing in \eqref{corrcorr1} are not equal to each other. However a generic statement cannot be made unless we 
actually solve all the EOMs. In view of that we will only demand \eqref{corrcorr1} as our constraint equation.  
 The other two equations involve relative signs and therefore 
tell us nothing about the signs of $\sum_{\{\alpha_i\} \ne 0} \mathcal{\tilde C}_{a} ^{a, \,\, i}$ or $\sum_{\{\alpha_i\} \ne 0}\mathcal{\tilde C }_{m} ^{m, \,\, i}$.
\newline
\newline
\noindent\fbox{%
    \parbox{\textwidth}{%
        In total we have the following conditions on the form of the corrections:
\begin{equation}
\label{overallconstraint} \frac{1}{2} \sum_{\{\alpha_i\}=0} \langle\mathcal{\tildeC}_{a}^{a, \,\, i}\rangle  
+ \frac{1}{4} \sum_{\{\alpha_i\}=0} \langle\mathcal{\tilde C}_{m}^{m, \,\, i}\rangle  <  \frac{2}{3}\sum_{\{\alpha_i\}= 0} \langle{\mathcal{\tilde C}}_{\mu} ^{\mu,\,\, i}\rangle ,
\end{equation}
\begin{equation}\label{corrcorr}
 \sum_{\{\alpha_i\} \ne 0} a^{\alpha_i} \langle{\mathcal{\tilde C}} _{\mu} ^{\mu, \,\, i}\rangle ~ >  ~ 0 .
\end{equation}
    }%
}

\section{Analysis of the background fluxes and additional consistency checks}
\label{fluxsxn}

{The above set of conclusions was derived by analyzing the Einstein's equations alone. The next question is whether any conclusions are altered when the equations of motion for the $G$-fluxes are taken into account.} Before moving ahead with the exact flux equations, we will do
a more careful analysis of the background fluxes to see how the type IIB fluxes should be viewed from our choices of the M-theory fluxes. Imagine we rewrite the flux components in M-theory
as \cite{dhhk}:
\bg\label{fluxu}
\widetilde{G} = G_{\mu\nu\rho m} dx^\mu\wedge dx^\nu \wedge dx^\rho \wedge dx^m + \widetilde{{\cal G}}_{mnqa}  dx^m \wedge dx^n \wedge dx^q \wedge dx^a
+ \sum_{i= 1}^N F^i \wedge \Omega^i ,\nonumber \\\nd
where we have taken the time-dependent components $\widetilde{G}_{mnpq}$ and $\widetilde{G}_{mnab}$ to be localized around certain singular points on the
eight-dimensional internal space
and we have decomposed $\widetilde{G}_{mnpa}$ into a {\it delocalized} and a {\it localized} piece as
\bg\label{locpiece} \widetilde{G}_{mnpa} = \widetilde{\cal G}_{mnpa} + \widetilde{G}_{mnpa}^{loc} .\nd
{In \eqref{fluxu}, the localized pieces are contained in the last term, where the sum is over the points at which the F-theory torus degenerates, the $\Omega^i$ are the normalizable harmonic forms located at these points, and the $F^i$ represent the gauge fields on the resulting D7-branes at these points in IIB, such that only the $F^i$ are functions of time.\footnote{A discussion of these issues is also given in \cite{becker1} and \cite{becker2}. Note 
that the existence of these points do not mean that the eight-dimensional manifold is singular.}}
Then it turns out that the delocalized piece $\widetilde{\cal G}_{mnpa}$ gives rise to the type IIB three-forms in the following way:
\bg\label{g2iib} 
\widetilde{\cal G}_{mnpa}
dx^m \wedge dx^n \wedge dx^p \wedge dx^a &\equiv & 2 (H_{3})_{mnp} dx^m \wedge dx^n \wedge dx^p \wedge dx^3\nonumber\\
&& + 2 (F_{3})_{mnp} dx^m \wedge dx^n \wedge dx^p \wedge dx^{11},
\nd
where $H_{3}$ and $F_{3}$ are  the NS and RR three-forms of type IIB theory respectively, while the localized
fluxes should appear as gauge-fields on the type IIB seven-branes.
A straightforward
decomposition immediately gives us:
\bg\label{m2b}
&&\int G\wedge \ast_{11} G ~  \rightarrow ~ \int d^{10}x \sqrt{g_{10}}~\left[
{1\over g_B^{2}} \left(\vert H_3 \vert^2 + \vert F_5\vert^2\right) + \vert F_3 \vert^2\right] \nonumber\\
&& ~~~~~~~~~~~~~~~~~~~~~~~~~ + \sum_{i = 1}^N~\int d^8 \sigma ~F^i \wedge \ast_B F^i , \nonumber\\
&& \int  C \wedge G \wedge G ~  \rightarrow ~ \int  ~ C_4 \wedge H_3
\wedge F_3 ~+~ \sum_{i = 1}^N \int d^8 \sigma ~C_4 \wedge F^i \wedge F^i , \nd
where for the first relation, the
first three terms appear in the type IIB bulk
and {the last term collects} the interactions
on the D7-brane worldvolume. We have also assumed that the self-duality of $F_5$ is imposed via the EOM, so that the action is explicitly non-selfdual. The five-form piece comes from the
spacetime part of the $G$-flux and the three-form fluxes come from the components $G_{mnqa}$. For the second relation, the first term is the bulk term and the second one is the
seven-brane Chern-Simons term.
The $C \wedge X_8$ term  gives rise to the couplings on the D7-branes
and O7-planes and possibly some contributions to the
bulk interactions. For example we expect some parts of $C \wedge X_8$ to reproduce
\bg\label{d7o7}
a_1\int_{\rm D7} C_{\rm RR} \wedge \sqrt{\hat{A}(R)} + a_2\int_{\rm O7} C_{\rm RR} \wedge \sqrt{H(R/4)} , \nd
where ${\hat{A}}(R)$ and ${H}(R)$ are the corresponding A-roof genus and Hirzebrusch polynomial respectively. We have also used the orthogonality condition for the components of $\Omega^i$ to get the interactions of the seven-brane worldvolume gauge fields. Note that this analysis only gives the abelian part of the gauge group (i.e the Cartan subalgebra), which could be extended to include a non-abelian gauge group by including M2-branes wrapping vanishing 2-cycles of the fourfold.


Once the structure of the fluxes is laid out, the physics away from the singular points will be captured by the delocalized fluxes only. 
The $G$-flux EOM \eqref{gfeom} then gives us the following equation for the warp factor $h$:\footnote{We have defined the covariant derivative $D_q$ in the following way: 
$D_q G^{qmnp} \equiv {1\over \sqrt{-g}} \partial_q\left(\sqrt{-g} G^{qmnp}\right)$.}
\bg\label{hosto} 
-\square h = && {1\over 12} {\widetilde {\cal G}}_{mnpa} (\ast_8{\widetilde {\cal G}})^{mnpa} 
+{2\kappa^2 T_2\over 8!\sqrt{\tilde g}} (X_8)_{M_1...M_8} {\widetilde \epsilon}^{M_1....M_8}\\
&& + {2\kappa^2 T_2 n_3 \over \sqrt{\tilde g}}\delta^8(x-X) - 
{2\kappa^2 T_2 {\bar n}_3 \over \sqrt{\tilde g}}\delta^8(x-Y) + \alpha_1 {\delta {\cal S}_{ntop}\over \delta \widetilde{C}_{012}} 
+ \alpha_2 {\delta {\cal S}_{top}\over \delta \widetilde{C}_{012}}, \nonumber \nd
where $\ast_8$ is the Hodge star with respect to the unwarped metric unless mentioned otherwise, $\alpha_i$ are coefficients that can be derived from \eqref{gfeom}, 
{and we take only the delocalized flux components.} Equation \eqref{hosto} can be compared to the Einstein equation:
\begin{eqnarray}\label{folrela}
- \Box h & = & \frac{\widetilde {\cal G}_{mnpa} \widetilde {\cal G}^{mnpa}}{12} + 12 h^2 \Lambda + \frac{2 \kappa^2 n_3 T_2 \delta^8(x - X)}{\sqrt{\tilde g}}
+ \frac{2 \kappa^2 \bar{n}_3 T_2 \delta^8(x - Y)}{\sqrt{\tilde g}}
\nonumber\\
&& + h^{4/3} \left (\frac{1}{2} \sum_{\{\alpha_i\}=0} \mathcal{\tildeC}_{a}^{a, \,\, i}  + \frac{1}{4} \sum_{\{\alpha_i\}=0} \mathcal{\tildeC}_{m}^{m, \,\, i} 
- \frac{2}{3}\sum_{\{\alpha_i\}= 0} {\mathcal{\tilde C}}_{\mu} ^{\mu,\,\, i}  \right),
\end{eqnarray}
where we have re-expressed \eqref{conoklo} in terms of the delocalized fluxes instead of the total fluxes.
The factors ($n_3, {\bar n}_3$) denote the number of M2 and anti-M2 branes located at ($X, Y$) respectively and $X_8$ is defined in the usual way \cite{Becker:1996gj} such that 
\bg\label{x8}
\int X_8 ~ = ~ -{1\over 4! (2\pi)^4} \chi_4 , \nd
where the integral is over the eight-dimensional manifold with Euler characteristic $\chi_4$, which could in general take any 
sign.

Comparing \eqref{folrela} and \eqref{hosto} we get the following consistency relation which should be compared with the consistency 
condition that we had from \eqref{conokl}:
\bg\label{consirel}
&& {1\over 12} {\widetilde {\cal G}}_{mnpa} \left[{\widetilde {\cal G}}^{mnpa} - (\ast_8{\widetilde {\cal G}})^{mnpa}\right] + 12 \Lambda h^2 + 
{4\kappa^2 T_2 {\bar n}_3 \over \sqrt{\tilde g}}\delta^8(x-Y) - \alpha_1 {\delta {\cal S}_{ntop}\over \delta \widetilde{C}_{012}}  
- \alpha_2 {\delta {\cal S}_{top}\over \delta \widetilde{C}_{012}} \nonumber\\
&& +  h^{4/3} \left (\frac{1}{2} \sum_{\{\alpha_i\}=0} \mathcal{\tildeC}_{a}^{a, \,\, i}  + \frac{1}{4} \sum_{\{\alpha_i\}=0} \mathcal{\tildeC}_{m}^{m, \,\, i}
- \frac{2}{3}\sum_{\{\alpha_i\}= 0} {\mathcal{\tilde C}}_{\mu} ^{\mu,\,\, i}  \right)
 - {2\kappa^2 T_2\over 8!\sqrt{\tilde g}} (X_8)_{M_1...M_8} {\widetilde \epsilon}^{M_1....M_8} = 0. \nonumber\\ \nd
Firstly note that in the presence of curvature corrections and positive cosmological constant $\Lambda$ it is in general {\it not} possible to maintain the self-duality of the 
$G$-fluxes. This may be more obvious if we re-express
 \eqref{threeeom} using \eqref{g2iib} as
\bg\label{3eomrex}
\vert H_{3} \vert^2 - \vert F_{3} \vert^2 = \frac{h^{4/3}}{12} \sum_{\{\alpha_i\}=0} \left(\mathcal{\tildeC}_{11}^{i} - \mathcal{\tildeC}_{33}^{i}\right),\nd 
which may not be consistent with $H_{3} = -\ast_6 F_{3}$ and $F_{3} = \ast_6 H_{3}$, where $\ast_6$ is the six-dimensional Hodge star measured with respect to the unwarped 
metric. 
In other words:
\bg\label{selfie} {\widetilde {\cal G}}^{mnpa} - (\ast_8{\widetilde {\cal G}})^{mnpa} \ne 0  ,\nd
meaning that supersymmetry should be broken to allow for a positive cosmological constant. One may also note that the contribution from the anti-M2 branes in \eqref{consirel} allows the self-duality of the G-fluxes to be broken even for vanishing cosmological constant $\Lambda$ and vanishing higher-order corrections. This means supersymmetry can be broken in {\it flat} space by the anti-M2 branes.

 The above relation can in fact be extended to the 
{\it full} G-fluxes, i.e. including both the localized and the delocalized pieces. 
{To show this we make use of another component of the $G$-flux equation, finding}
\bg\label{selfie2} 
\Lambda(t) D_q {\widetilde G}^{qmnp} +  D_a {\widetilde G}^{amnp} &=& 
{\partial_q h\over h}\left[\Lambda(t) {\widetilde G}^{qmnp} - {1\over 12}\left(\ast_8{\widetilde G}\right)^{qmnp}\right] \\ 
&& + {\partial_a h\over h}\left[{\widetilde G}^{amnp} - {1\over 12}\left(\ast_8{\widetilde G}\right)^{amnp}\right] 
+ \beta_1 {\delta {\cal S}_{ntop}\over \delta \widetilde{C}_{mnp}} + \beta_2 {\delta {\cal S}_{top}\over \delta \widetilde{C}_{mnp}}, \nonumber \nd
which is expressed in terms of the total fluxes and is 
again consistent with \eqref{selfie}. In deriving the above equation we have assumed 
\bg\label{assumed}
\left(X_8\right)_{012M_1....M_5} \approx 0.\nd
Note that for the delocalized flux components $\widetilde{\cal G}_{mnpa}$, away from the singular points, \eqref{selfie2} simplifies to 
\bg\label{selfsim} 
D_a {\widetilde {\cal G}}^{amnp} = {\partial_a h\over h}\left[{\widetilde {\cal G}}^{amnp} - {1\over 12}\left(\ast_8{\widetilde {\cal G}}\right)^{amnp}\right]
+ \left[\beta_1 {\delta {\cal S}_{ntop}\over \delta \widetilde{C}_{mnp}} 
+ \beta_2 {\delta {\cal S}_{top}\over \delta \widetilde{C}_{mnp}}\right]_{\widetilde{G}^{loc}_{mnpa}= 0}^{\widetilde{G}_{mnpq} = 0}\nd
meaning that the delocalized flux components are not covariantly constant. Another consequence of the above equation is that the $\widetilde{\cal G}_{mnpa}$ components will continue to remain time independent provided 
\bg\label{gutcons}
{\partial\over \partial t}\left[\beta_1 {\delta {\cal S}_{ntop}\over \delta \widetilde{C}_{mnp}} 
+ \beta_2 {\delta {\cal S}_{top}\over \delta \widetilde{C}_{mnp}}\right]_{\widetilde{G}^{loc}_{mnpa}= 0}^{\widetilde{G}_{mnpq} = 0} ~ = ~ 0, \nd 
giving us another constraint on the curvature corrections in the theory, although solutions should also exist for cases which violate this constraint and hence require a more general analysis that includes a time dependence for $\widetilde{\cal G}_{mnpa}$. 

Now looking at \eqref{selfie} and \eqref{hosto} we conclude that a four-fold with {\it negative} Euler characteristic $\chi_4$ may easily accommodate fluxes of the kind \eqref{selfie} and simultaneously account for the supersymmetry breaking, although this is not a necessary condition for a solution to exist. In other words, without loss 
of generality, we can demand
\bg\label{demand} {1\over 4}\int \sqrt{\tilde g}~ \widetilde{{\cal G}}_{mnpa} \left(\ast_8 \widetilde{{\cal G}}\right)^{mnpa} ~=~ \int H_{3} \wedge F_{3} ~ < 0, \nd 
which in turn can be made consistent with the first equation in \eqref{threeeom}, namely
\bg\label{13eom}
\int d^6 x \sqrt{\tilde g} ~(H_{3})_{mnp} (F_{3})^{mnp} = -  3 \sum_{\alpha_i = 0} \langle {\cal C}^i_{3, 11} \rangle ,  \nd
provided $\sum_{\alpha_i = 0}\langle {\cal C}^i_{3, 11}\rangle > 0$. This could be taken as another constraint on the curvature corrections, 
which applies in the case that $\chi_4<0$. A similar constraint would apply for the case $\chi_4 >0$.


{Yet another possible class of solutions are those with vanishing Euler characteristic $\chi_4=0$. These solutions could correspond to an internal M-theory eight manifold that is an elliptical fibration of a Calabi-Yau threefold, since the Euler characteristic of the eight manifold is related to the Chern classes of the base by \cite{oai:arXiv.org:hep-th/9606122}}:
\begin{equation}
\chi_4 = 12 \int_B c_1 (c_2 + 30 c_1 ^2).
\end{equation}
If the base manifold is Calabi-Yau, then $c_1=0$, and hence $\chi_4$ vanishes. This, in conjunction with the condition $\tilde{R}=0$, leads to its own set of solutions, with the modified conditions:
\begin{eqnarray}
\sum_{\{\alpha_i\}=0} \langle \mathcal{\tildeC}_{m}^{m, \,\, i} \rangle_{\chi=0} & < & 0, \\
\sum_{\{\alpha_i\}=0} \langle \mathcal{\tildeC}_{a}^{a, \,\, i} \rangle _{\chi=0}& < &0.
\end{eqnarray} 
As an interesting corollary, in the {\it absence} of any curvature corrections and due to \eqref{secondconst}, \eqref{thirdconst} or \eqref{jom}, it is {\it impossible} to
get a four-dimensional de Sitter spacetime if the internal six-dimensional base of the M-theory eight-fold is a Calabi-Yau manifold because
\bg\label{colbi}
\tilde{R} = -18 h \Lambda. \nd
We now make a few observations. Note that to stabilize all the complex structure moduli, we will have to switch on $G$-fluxes in the internal manifold. 
The $\widetilde{{\cal G}}_{mnqa}$ 
components are the ones that will do the 
required job for us. However due to the background constraint \eqref{selfie} we cannot allow supersymmetric fluxes. In fact we can extend \eqref{selfie}, by incorporating 
the localized fluxes in \eqref{hosto} and \eqref{folrela}, to {\it full} G-fluxes $G_{mnpa}, G_{mnpq}$ and $G_{mnab}$. This means, in addition to \eqref{selfie} we will have
another relation
\bg\label{selFF}
G^{loc} - \ast_4 G^{loc} \ne 0, \nd  
where $\ast_4$ is the Hodge star on a four-dimensional surface $\Sigma_4$ inside the six-dimensional base of our eight-manifold. Since the localized fluxes are related 
to the gauge fields on the seven-branes wrapping $\Sigma_4$ in type IIB theory, 
this immediately implies that the gauge fluxes (both the abelian and the non-abelian pieces)
will create a D-term potential satisfying the background constraint relations \eqref{consirel} and \eqref{conokl}. 

In addition to that, the decomposition \eqref{locpiece} switches on an FI term from the $H_3 = dB_2$ of $\widetilde{{\cal G}}_{mnqa}$ and from the $F_2 = dA$ of 
$\widetilde{G}^{loc}_{mnqa}$, proportional to
\bg\label{fiterm}
\int_{\Sigma_4} {\cal F}^- \wedge {\cal F}^-  \nd
where ${\cal F}^- \equiv {\cal F} - \ast_4 {\cal F}$ and we have defined  ${\cal F} \equiv F_2 - B_2$. 
   
Since the background supersymmetry is broken by the $G$-fluxes, the 
F-term is explicitly non-zero allowing us to switch on a non-zero D-term in the presence of higher-curvature quantum corrections. 
The fact that the F-term and D-term are related to each other can be inferred from the decomposition \eqref{locpiece} where {\it both} three-form and gauge
fluxes in type IIB are sourced by M-theory
G-fluxes. This way we take care of the issues raised by \cite{nilles}.\footnote{It will be interesting to compare our results with the ones in \cite{achucarro} regarding
D-term uplifting.}  
Note that in the {\it absence} of the quantum corrections, this
wouldn't have been possible. 

Finally, we need to switch on D-brane instantons that would help us stabilize all the K\"ahler structure moduli, including the volume moduli. As mentioned earlier, we have
to make sure that the internal manifold is stabilized at large volume so that the dynamics can be captured by the set of EOMs described above. 
In the presence of the D-brane instantons 
higher-curvature terms are automatically generated (some aspect of this will be discussed in Section \ref{curvaturedisc}). These curvature terms are the last pieces of the link required to satisfy the consistency
relations \eqref{conokl} or \eqref{consirel}.  

Thus both the fluxes and the curvature corrections are therefore {\it necessary} consequences of stabilized moduli in this set-up. As such they could lead to a positive cosmological constant solution, and a natural realization of D-term uplifting \cite{Burgess:2003ic}.

\section{A discussion on the curvature corrections}
\label{curvaturedisc}


In this section we discuss in more detail the possible origins for the higher-order curvature corrections\footnote{We will restrict ourselves to $R^n$ corrections as these have been 
studied in more detail than the $G^n$ corrections. For an analysis of $G^n$ corrections, the readers may refer to \cite{deser, jock}.} 
we have argued might allow for construction of de Sitter vacua in IIB compactifications. While our calculations were done in M-theory, it is interesting to first look at the corrections that can appear in type IIB string theory. These terms can be sourced by tree- and loop-level $n$-graviton scattering amplitudes, or equivalently loop corrections to the underlying $\sigma$-model, and are also induced by D-instanton corrections. The general form of these corrections is given by (adopting the notation of \cite{Stieberger:2009rr}, combined with \cite{Green:2005ba} but with the substitution $s=(m+6)/4 $):
\begin{equation}
(\alpha')^{n-m+1} t_{m,n}  Z_{m} ^{(w,w')} D^{2m} R^n
\end{equation}
where $ t_{m,n} D^{2m} R^n$ is the contraction of $2m$ covariant derivatives and $n$ Riemann tensors with a tensor $t_{m,n}$. The coefficient $Z^{w,w'} _{m}$ is an eigenfunction of the Laplace operator on the fundamental domain of $SL(2,\mathbb{Z})$, with modular weight $(w,w')$. This coefficient can be written as an Eisenstein series \cite{Green:2005ba}, and is
 necessary for $SL(2,\mathbb{Z})$ invariance of the corrections to the action.

The lowest-order correction can be calculated from $4$-graviton scattering; see for example \cite{Gross:1986iv} in type II and \cite{Gross:1986mw} in Heterotic, which induces a $D^0\
 R^4$ correction at both tree level (at order $(\alpha')^3$ ) and at the one-loop level. In the calculation by Gross and Witten \cite{Gross:1986mw}, this led to a gaussian path integral that can alternatively be written as a contraction of four copies of the Riemann tensor with two copies of a rank-8 tensor denoted $t_8$. This allows one to write the correction as (equations 10 and 11 of Gross and Witten):
\begin{equation}
 \int \mathrm{d} \psi^{\alpha}_{L} \mathrm{d} \psi_{R} ^{\beta}\; \mbox{exp} \left[ \bar{\psi}^{\alpha} _L \Gamma^{\mu \nu} _{\alpha \beta} 
\psi^\beta _L \bar{\psi} ^{\alpha'} _R  \Gamma^{\sigma \tau} _{\alpha' \beta'} \psi^{\beta'} _R R_{\mu \nu \sigma \tau}\right],
\end{equation}
or in terms of the $t_8$ tensor:
\begin{equation}
t^{\mu_1 \mu_2 ... \mu_8} t^{\nu_1 \nu_2 ... \nu_8} R_{\mu_1 \mu_2 \nu_1 \nu_2} R_{\mu_3 \mu_4 \nu_3 \nu_4} R_{\mu_5 \mu_6 \nu_5 \nu_6} 
R_{\mu_7 \mu_8 \nu_7 \nu_8},
\end{equation}
with the $t_8$ tensor defined by
\begin{equation}
\sqrt{\mbox{det} \Gamma^{\mu \nu} F_{\mu \nu}} = t^{\mu_1 \mu_2 ... \mu_8} F_{\mu_1 \mu_2} F_{\mu_3 \mu_4} ... F_{\mu_7 \mu_8} .
\end{equation}

The above correction is often written in the literature as simply $t_8 t_8 R^4$. Another approach to calculating this correction is to consider loop corrections in the sigma model (see for example \cite{Grisaru:1986vi}), where an $n$-loop effect will lead to an $R^n$ correction that is order $(\alpha')^n$ in the corresponding string theory. Collecting all the terms at order $R^4$ yields a correction of the form:
\begin{equation}
\left(\frac{1}{8}\epsilon_{10} \epsilon_{10}  - t_8 t_8\right)R^4,
\end{equation}
where $\epsilon_{10}$ is the rank-10 totally anti-symmetric tensor .

One might also wonder if there are $R^2$ or $R^3$ terms. The sigma model analysis does not produce these terms, which would indicate that type II theories are protected from $\alpha'^2$ and $\alpha^3$ corrections, as shown in the sigma model in \cite{Grisaru:1986px}. This was also done in the context of type I, II and heterotic string theory in \cite{Metsaev:1986yb}, which confirmed the result that $R^2$ and $R^3$ corrections do not appear. One can also check that $R^5$ terms do not arise, and in fact the next corrections coming from the tree-level graviton scattering are $D^2 R^4$, $D^2 R^5$, and $R^6$, all at order $(\alpha')^5$ (see table I of \cite{Stieberger:2009rr}). At the loop level, there has been recent work \cite{Berg:2007wt, Cicoli:2007xp, Pedro:2013qga} showing that perhaps string loop corrections at order $g_s ^2 (\alpha')^2$ can become important in a certain class of compactifications (dubbed the Large Volume Scenario).

Another contribution comes from calculating the graviton scattering amplitude in a D-instanton background, as was done by Green and Guterperle \cite{Green:1997tv}, which gives an extra contribution to $Z_{m} ^{(w,w')}$ that is neccessary for the correction to be $SL(2,\mathbb{Z})$ invariant. The coefficient for the $D^0 R^4$ correction has modular weight $(w,w')=(0,0)$, and  is given by (equation 1.15 of \cite{Green:2005ba} with $ s=3/2$, or in our notation, $m=0$):
\begin{eqnarray}
Z_0 = && 2 \zeta(3) {C^{(0)}} ^{3/2} + 8 \zeta(3) {C^{(0)}}^{-1/2} \\
\nonumber && + 4 \pi \displaystyle \sum_{k \neq 0} \mu(k,3/2) \exp \left[- 2\pi(|k| e^{- \phi} - i k e^{- \phi}) \right] 
\sqrt{|k|} \left( 1+ \frac{3}{16 \pi |k| C^{(0)}}+ ...\right) ,
\end{eqnarray}
where $C^{(0)}$ and $\phi$ are the axion and dilaton. The first term on the RHS is the tree-level correction, while the second term is the 1-loop correction. The set of terms on the \
second line is an infinite set of D-instanton corrections, with the function $\mu(k,3/2)$ defined as in Appendix A of \cite{Green:2005ba}.

The picture in M-theory is slightly simpler, as there is only one curvature superinvariant. A review of the corrections to M-theory supergravity, as well as the supersymmetrization, can be found in \cite{Howe:2004pn}, while the detailed derivations  can be found in \cite{Howe:2003cy} and \cite{Anguelova:2004pg}.  A feature of the M-theory picture that is fairly 
well understood is the necessity of an additional Chern-Simons term to cancel the 5-brane anomaly, via anomaly inflow. This term takes the form
\begin{equation}
C \wedge X_8 ,
\end{equation}
where $X_8$ is built out of $R^4$. As this term includes a factor of the M-theory $3$-form flux, it will contribute to the equation of motion of the fluxes. 

A key feature of these corrections is that the form of the contraction conspires to choose only the Weyl part of the Riemann tensor, such that the corrections vanish on manifolds with vanishing Weyl tensor. This was shown explicitly by Banks and Green in \cite{Banks:1998nr}, where they considered $AdS_5 \times S^5$. This is great news for AdS/CFT, since the correspondence is protected from loop corrections.  However, it makes the search for scenarios where corrections may be important a non-trivial exercise. One possibility for finding non-negligible corrections is to consider Calabi-Yau manifolds, and indeed this is the internal manifold used in the 4D effective picture of these corrections in Kahler Uplifting \cite{Westphal:2006tn, Louis:2012nb}. However, this introduces a new difficulty: many Calabi-Yau manifolds can not be given an explicit metric -- for example the explicit realization of Kahler uplifting in \cite{Louis:2012nb} is done on $\mathbb{CP}^{11169}$.


\section{Conclusion}
This paper has been a close examination of de Sitter solutions in Type IIB string theory, from the perspective of the 10-dimensional equations of motion (and the corresponding 11-dimensional M-theory equations). We have reached two key conclusions: 

\begin{enumerate}
\item  By applying the Gibbons-Maldacena-Nunez No-Go Theorem \cite{Gibbons:1984kp, Gibbons:2003gb, Maldacena:2000mw} to localized static sources we have found that the inclusion in IIB supergravity of Dp- branes, anti Dp-branes, Op-planes, 
and by extension any linear combination thereof, does not lead to  positive curvature in the 3+1 non-compact directions. \\
\item The addition of curvature corrections, sourced by D-instantons as well as tree and loop-level graviton scattering, may lead to a de Sitter solution in the 3+1 non-compact directions, although an explicit construction of this would require specifying a metric on the internal manifold as well as a subset of correction terms to consider. Furthermore, this solution
naturally leads to compactification with broken supersymmetry, all moduli stabilized, and the generation of a D-term in the scalar potential of the 4d effective field theory.
\end{enumerate}

The first result is a fairly simple extension of the analysis performed by Maldacena-Nunez \cite{Maldacena:2000mw}, and Giddings, Kachru, Polchinski \cite{kachruone}, among others. Our assumptions in deriving this were limited to demanding (i) maximal symmetry  in the 3+1 dimensions, as well as (ii) positive curvature in the 3+1 dimensions. Since we only consider time-independent matter configurations, the 3+1 dimensional non-compact spacetime we are looking for is  `pure' de Sitter, as opposed to quasi-de  Sitter as is usually considered in cosmology. However, to construct any 3+1 dimensional positive curvature geometry, the  stress-energy tensor must satisfy the condition (2.8) regardless of the symmetry, and in particular regardless of time dependence.

Note that there are many existing proposals which we have not considered, for example IIA on nilmanifolds \cite{Silverstein:2007ac}, IIA on solvmanifolds \cite{Andriot:2010ju}, and non-geometric fluxes \cite{Danielsson:2012by}. These proposals should also be subject to condition (2.8).

The second result is a non-trivial check that curvature corrections do indeed evade the No-Go theorems. In this calculation we have used an ansatz for the effective stress-energy tensor induced by the curvature corrections, which we view as an appropriate way to proceed given the freedom to set the internal manifold as well as the complicated (and not completely known) form of the curvature corrections. 

A worthy question at this point would be the sensitivity of our second result to the form of the ansatz, as it is entirely possible that some choices of internal manifold do not lead to curvature corrections that can be parametrized in this way. Thus a conservative restatement of our second result would be as follows: given a class of internal manifolds that allow the time dependence of the curvature correction to be isolated from other contributions, there do exist de Sitter solutions provided a set of consistency conditions \eqref{overallconstraint} - \eqref{corrcorr} is satisfied. This hints at interesting further work, to clarify the consistency of our claims with the work of Sethi et al. \cite{Green:2011cn} which found that such corrections in Heterotic theory do \emph{not} lead to dS solutions.

Upon studying the dS solution obtained via curvature corrections, we uncovered a number of interesting features. Solutions exist for any choice of the Euler characteristic of the internal manifold, including an elliptic fibration of a Calabi-Yau threefold. Furthermore, this setup generically leads to non self-dual fluxes, which break supersymmetry, and induce a D-term in the scalar potential, suggesting that this construction may be a realization of D-term uplifting \cite{Burgess:2003ic}. The moduli of this setup can be fully stabilized: the complex structure moduli are fixed by the fluxes, while the K\"{a}hler moduli are stabilized by the D-instantons, which in turn source the curvature corrections. Hence our analysis indicates that curvature corrections \emph{can} do the job at hand.

This work has opened up several directions for future research. One option, motivated by the desire for a deeper understanding of string theory, is to continue the investigation of de Sitter solutions, using dualities to relate the solutions in different string theories. This has the potential to clarify subtleties of dualizing non-BPS states, and to allow one to `map out' the space of dS vacua in string theory.

An alternative way forward is to push this work closer to cosmology, and in particular, inflationary cosmology. While the full 10d equations do not lend themselves to model building, this approach does provide a clear path to studying compactifications with a time-dependent scalar curvature (`quasi-dS').  The appeal of this option lies in building self-consistent embeddings of inflationary cosmology in string theory, with the (albeit ambitious) goal of teasing out distinctive signatures of string theory in the sky. As has happened before, it may be that effects from a full 10-dimensional construction result in observational signatures which do not arise in the effective field theory approach.

\section*{Acknowledgements}
We would like to thank Juan Maldacena, Martin Kruczenski, Peter Ouyang, Savdeep Sethi, Sunil Mukhi, Daniel Elander, Anatoly Dymarsky, Ana Achucarro, Johannes Walcher, Sergei Khlebnikov  and Shamit Kachru for helpful discussions, as well as the Banff International Research Station where a portion of this work was completed. KD and EM thank Robert H. Brandenberger for many very helpful discussions during the completion of this work. The work of KD is supported in part by the National Science and Engineering Research Council of Canada. RG is grateful for the support of the European Research Council via the Starting Grant Nr. 256994 ``StringCosmOS''. The work of MM is supported by DOE through grant DE-SC0007884. We thank Lucie Baumont for her work in the initial stages of this project, and Daniel Junghans for insightful comments on the first draft of this paper.

\newpage


{}
\end{document}